\documentclass{article}
\usepackage[dblblindworkshop,final,nonatbib]{neurips_2026}
\workshoptitle{Sim2Science: ML with Imperfect Scientific Models}
\usepackage[numbers,sort&compress]{natbib}

\usepackage[utf8]{inputenc}
\usepackage[T1]{fontenc}
\DeclareUnicodeCharacter{2009}{\,}
\DeclareUnicodeCharacter{2248}{\ensuremath{\approx}}
\usepackage{amsmath}
\usepackage{amsfonts}
\usepackage{amssymb}
\usepackage{booktabs}
\usepackage{float}
\usepackage{graphicx}
\usepackage{microtype}
\usepackage{nicefrac}
\usepackage{xcolor}
\usepackage{url}
\usepackage{hyperref}

\graphicspath{{figures/}}

\newcommand{\aff}[1]{\textsuperscript{\normalfont #1}}
\title{A Cross-Domain Simulator and Benchmark for Uncertainty Quantification in Global 21-cm Inference}
\author{%
  \textbf{Jacob Tutt}\aff{1} \quad \textbf{Harry Bevins}\aff{1,2} \quad \textbf{James Alvey}\aff{1} \quad \textbf{Dominic Anstey}\aff{1} \\
  \textbf{John Cumner}\aff{1} \quad \textbf{Eloy de Lera Acedo}\aff{1} \quad \textbf{Anastasia Fialkov}\aff{1} \\[0.5em]
  \aff{1}University of Cambridge, Cambridge CB3 0HA, UK \\
  \aff{2}Imperial College London, London SW7 2AZ, UK \\
  \texttt{jlt67@cam.ac.uk}
}

\begin{document}

\maketitle

\begin{abstract}
End-to-end uncertainty propagation in scientific inference often requires repeated evaluation of coupled simulators spanning multiple physical domains, often at prohibitive computational cost. Global 21-cm cosmology is an extreme example: a millikelvin cosmological signal must be separated from radio foregrounds and instrumental effects four orders of magnitude larger, making inference acutely sensitive to model misspecification. Reliable recovery is therefore contingent on an exceptionally accurate model that jointly propagates uncertainty through the complete observational chain. We present a differentiable, GPU-accelerated hybrid forward model combining neural emulation of the cosmological signal, a physics-informed multi-resolution parameterisation of the galactic foregrounds, and a linear surrogate model of the antenna. This unifies simulators spanning astrophysics and computational electromagnetics while reducing evaluation time by a factor of $7.2\times10^{7}$. Using GPU-accelerated Nested Slice Sampling, we benchmark the framework across 100 unseen signal realisations and find that 94 injected profiles lie entirely within their corresponding 95\% credible bands. Together, the simulator and reference Nested Slice Sampling results provide a demanding benchmark of posterior calibration, mode coverage, and evidence accuracy across next-generation stochastic samplers, simulation-based inference, and variational inference frameworks.
\end{abstract}

\section{Introduction}
\label{sec:introduction}

Observations of the first billion years of cosmic history remain sparse and indirect, leaving the astrophysical processes governing the formation of the first stars and galaxies weakly constrained by existing data. By tracing the thermal and ionisation state of the intergalactic medium (IGM), the redshifted 21-cm hyperfine transition of neutral hydrogen offers one of the most direct probes of this period \citep{Furlanetto_2006,Pritchard_2012}. Single-antenna radiometer experiments \citep{Bowman_2018,Philip_2018,Singh_2018,Monsalve_2024,Bull_2025,Acedo_2022} target its sky-averaged brightness temperature, $T_{21}(\nu)$, but realising their potential remains a formidable inference challenge \citep{Harker_2012,Liu_2013,Switzer_2014,Monsalve_2017,Tauscher_2018a,Tauscher_2018b,Hills_2018,Singh_2019,Sims_2020,Rapetti_2020,Tauscher_2020a,Tauscher_2020b,Tauscher_2021,Bevins_2021,Anstey_2022,Murray_2022,Saxena_2023,Hibbard_2023,saxena_2024,Sims_2025,sims_2023,Tutt_2026a}.

Theoretical models predict the global signal to be of order a few hundred millikelvin \citep{Fialkov_2014}, yet it must be extracted from Galactic and extragalactic foreground emission four orders of magnitude brighter \citep{Shaver_1999}. Moreover, the antenna's direction- and frequency-dependent response to the radio sky, itself highly sensitive to geometric and environmental perturbations, non-trivially couples spatial foreground structure into the observed spectrum, producing features that are highly degenerate with the cosmological signal \citep{Anstey_2021}. Crucially, neither the foreground nor instrumental contributions can be characterised \textit{a priori} to the required precision, and modelling deficiencies in either can leave residuals far exceeding the signal itself \citep{Hibbard_2020,Bassett_2021a,Cumner_2024,Pattison_2026}. Global 21-cm inference therefore operates in a severe model-misspecification regime, in which an analysis may appear well constrained while recovering a systematically biased signal and a severely miscalibrated posterior.

Rigorous inference consequently requires a forward model that propagates uncertainty across the observational chain by coupling cross-domain simulators, from semi-numerical astrophysical simulations \citep{Cohen_2017} to computational electromagnetic (CEM) antenna solvers \citep{Harrington1993}. The dimensionality of this uncertainty space compounds the cost of constructing large ensembles for simulation-based inference (SBI) \citep{Cranmer_2020} or repeatedly evaluating the model within Bayesian samplers. Forward-model acceleration addresses only one part of the computational challenge. The large dynamic range and strong degeneracies produce correlated, multimodal posterior geometries, posing a stringent test of sampling performance. These challenges motivate our accelerator-optimised hybrid framework. Implemented in JAX and coupled to GPU-native nested slice sampling (NSS) \citep{yallup2025nested,cabezas2024blackjax}, it enables end-to-end Bayesian inference over the global 21-cm observational chain at tractable computational cost.

Beyond its scientific application, this work addresses a timely methodological need: benchmarks must evolve with increasingly complex, high-dimensional scientific inverse problems. This 137-parameter problem provides a demanding open-source benchmark\footnote{The code required to run the benchmark is publicly available on \href{https://github.com/JacobTutt/global-21cm-inference-benchmark}{GitHub}.}, with reference posterior samples and evidence estimates for evaluating next-generation Monte Carlo, simulation-based, and variational inference methods. The accelerated surrogate model is presented in Section~\ref{sec:framework} and statistically benchmarked in Section~\ref{sec:experiments}, with limitations and future work discussed in Section~\ref{sec:limitations}.

\section{Hybrid Accelerated Forward Model}
\label{sec:framework}

\begin{figure}[!b]
    \centering
    \includegraphics[width=\textwidth]{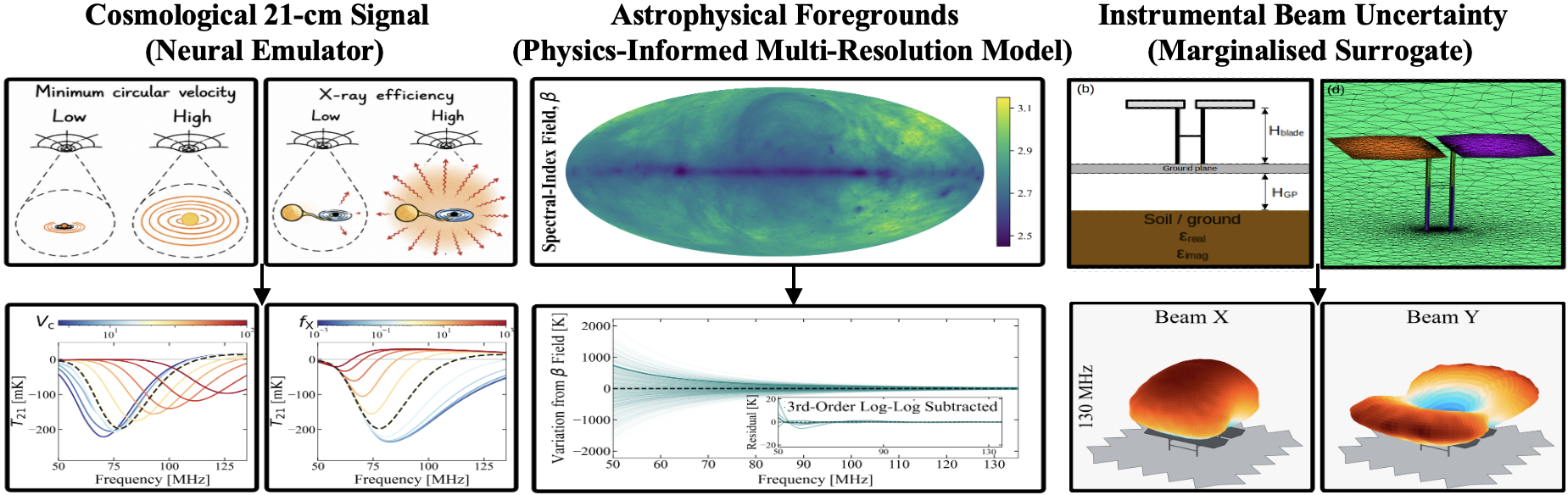}
    \caption{\textbf{Hybrid forward-model architecture.} Top: physical sources of uncertainty; bottom: their effect on the model. Left: a neural emulator maps astrophysical parameters to the global 21-cm signal. Centre: a physics-informed multi-resolution model represents uncertainty in the foreground sky. Right: a linear surrogate reconstructs chromatic beam variations from antenna and soil perturbations.}
    \label{fig:model_components}
\end{figure}

Figure~\ref{fig:model_components} summarises the three primary uncertainty contributions required to construct physically informed global 21-cm observations. Even when generated concurrently, a full-fidelity CEM beam ($\sim1$ hour across 1,600 CPU cores) and astrophysical signal (2.8 hours across three cores) require $\sim2.8$ hours of wall time and 1,608 CPU-core hours (\pounds16.08) per realisation.\footnote{Benchmarked using Intel Xeon Platinum 8368Q processors with 6.76~GB of RAM per core. Costs use the current \href{https://www.csd3.cam.ac.uk/charges}{CSD3 rate} of \pounds0.01 per CPU-core hour.} Consequently, conventional analyses often simplify or fix poorly constrained components of the observational chain. We instead present a framework that accelerates each component independently, with their unified implementation and resulting computational acceleration quantified in Section~\ref{sec:inference}.

\subsection{Neural emulation of the global 21-cm signal}
\label{sec:signal_emulator}

Simulating the global 21-cm signal requires the modelling of the radiative coupling, heating, and ionisation of the IGM and hence consideration of large-scale density and radiation fields alongside sub-grid prescriptions for halo and star formation. We simulate these processes with the semi-numerical code \texttt{21cmSPACE} \citep{Fialkov_2014,Cohen_2017}, adopting the model of \citet{Reis_2020_radio}, parameterised by six astrophysical quantities, $\boldsymbol{\theta}_{21}=\{f_{\star,\mathrm{II}},f_{\star,\mathrm{III}},V_{\mathrm{c}},f_{\mathrm{X}},\tau,f_{\mathrm{radio}}\}$ (see Appendix~\ref{app:signal_physics} for physical prescriptions).

To enable repeated evaluation during inference, we construct a neural emulator following \citet{Bevins_2021_globalemu}. Rather than predicting a fixed discretised spectrum, we train it on \texttt{21cmSPACE} realisations with redshift supplied as a variable alongside the astrophysical parameters. This allows the nonlinear evolution of the global 21-cm signal to be represented by a compact network, while complete spectra across all redshifts can be evaluated in parallel. Across the test set, it achieves a RMSE of $0.91~\mathrm{mK}$, more than an order of magnitude below the assumed $25~\mathrm{mK}$ instrumental-noise level and sufficiently low to avoid appreciable bias in downstream inference \citep{Bevins_2025}. Appendix~\ref{app:signal_emulator} specifies the dataset split, network architecture, training procedure, and evaluation underlying these benchmarks.

\subsection{Physics-informed Foreground Model}
\label{sec:foregrounds}

Although the low-frequency radio sky is uncertain in both its absolute brightness and spectral structure, its broad spatial structure is well constrained by existing radio surveys and provides an informative physical template. We therefore parameterise regional departures from the Global Sky Model (GSM) reference map at $\nu_0=230~\mathrm{MHz}$, $T_{230}(\Omega')$, in celestial coordinates $\Omega'$ \citep{deOliveiraCosta2008}, using independent sky partitions: $N_{\alpha}$ base-map correction regions with parameters $\alpha_i$ \citep{Pagano_2023}, and $N_{\beta}$ spectral-index regions with parameters $\beta_j$ \citep{Anstey_2021}, collectively $\boldsymbol{\theta}_{\mathrm{FG}}\equiv\{\boldsymbol{\alpha},\boldsymbol{\beta}\}$. The regional construction and prior specification are provided in Appendix~\ref{app:foreground_model}. For given binary mask $M_{i,j}(\Omega')$ defining the intersection of the $i$-th base-map and $j$-th spectral-index region, the sky model is then:
\begin{equation}
T_{\mathrm{sky}}^{\mathrm{model}}(\Omega',\nu;\boldsymbol{\alpha},\boldsymbol{\beta})
=
\left[T_{230}(\Omega')-T_{\mathrm{CMB}}\right]
\sum_{i=1}^{N_{\alpha}}\sum_{j=1}^{N_{\beta}}
M_{i,j}(\Omega')\alpha_i
\left(\frac{\nu}{\nu_0}\right)^{-\beta_j}
+T_{\mathrm{CMB}},
\label{eq:regional_sky_model}
\end{equation}
where $T_{\mathrm{CMB}}$ is the cosmic microwave background temperature.

\subsection{Marginalised surrogate of beam chromaticity}
\label{sec:beam_surrogate}
High-fidelity CEM solvers \citep{Cavillot_2021,Hu1999,Alkhalifeh2016,Cavillot2020,Cavillot2024} map a given antenna geometry and environmental conditions to a radiation pattern by solving the Electric Field Integral Equations. Since these physical parameters are purely instrumental nuisance quantities, we instead learn their induced beam uncertainty directly from 400 CEM realisations \citep{Tutt_2026b}. The surrogate must span the full beam-uncertainty space, encompassing variations of tens of kelvin, while retaining sensitivity to the millikelvin cosmological signal. We therefore introduce a hierarchical two-stage decomposition \citep{Tutt_2026b}. An initial singular value decomposition (SVD) extracts shared angular basis functions, $B_k(\Omega)$ \citep{Pieterse_2024}, while a second decomposition captures correlations in their weights across frequency and angular mode, compressing the representation to just $M=100$ coefficients. The resulting surrogate achieves a mean antenna-temperature reconstruction RMSE of $24.47~\mathrm{mK}$, below the instrumental-noise level. The beam model can be written as

\begin{equation}
\tilde{D}(\nu,\Omega;\boldsymbol{a})
=
\sum_{k=1}^{R}
\left[
e_{0,k}(\nu)
+
\sum_{m=1}^{M}a_m e_{m,k}(\nu)
\right]
B_k(\Omega),
\label{eq:beam_surrogate}
\end{equation}

where $\Omega$ denotes the antenna frame, $R$ the number of retained angular modes, $e_{0,k}(\nu)$ the mean spectral weight, and $e_{m,k}(\nu)$ the correlated correction modes, weighted by the beam coefficients $\boldsymbol{\theta}_{\mathrm{Beam}}\equiv\boldsymbol{a}=\{a_m\}$. The CEM training set, surrogate construction, and validation are documented in Appendix~\ref{app:beam_surrogate}.

\subsection{Accelerator-native implementation and performance}
\label{sec:inference}
\begin{figure}[!t]
    \centering
    \includegraphics[width=\textwidth]{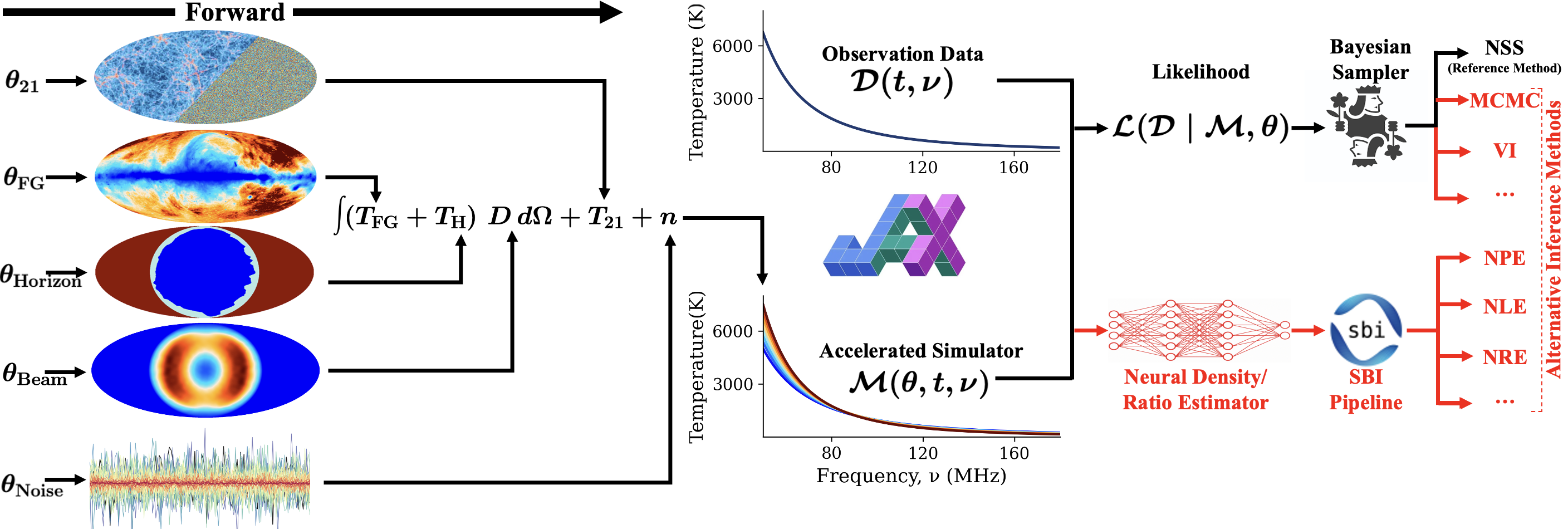}
    \caption{\textbf{Benchmark inference architecture.} Integration of full-fidelity observations and the accelerated simulator with the reference NSS analysis (black) and alternative inference methods (red).}
    \label{fig:pipeline_overview}
\end{figure}

To maximise computational throughput, we precompute the time- and frequency-dependent coupling between foreground regions and beam modes for the fixed observing configuration as a reusable response operator:
\begin{equation}
\mathcal{B}_{i,j,m}(\nu,t)
=
\frac{1}{4\pi}
\int_{4\pi}
M_{i,j}(\Omega,t)
\left[T_{230}(\Omega,t)-T_{\mathrm{CMB}}\right]
\sum_{k=1}^{R}
e_{m,k}(\nu)B_k(\Omega)
\,\mathrm{d}\Omega ,
\label{eq:basis_response}
\end{equation}
where $M_{i,j}$ and $T_{230}$ are rotated into the antenna frame at observing time $t$. The model antenna temperature, $T_{\mathrm{m}}$, then becomes (Figure~\ref{fig:pipeline_overview})
\begin{equation}
T_{\mathrm{m}}(\nu,t)
=
\sum_{m=0}^{M}a_m
\sum_{i=1}^{N_{\alpha}}
\sum_{j=1}^{N_{\beta}}
\alpha_i\mathcal{B}_{i,j,m}(\nu,t)
\left(\frac{\nu}{\nu_0}\right)^{-\beta_j}
+T_{\mathrm{CMB}}
+T_{21}(\nu;\boldsymbol{\theta}_{21})
+n(\nu,t;\boldsymbol{\theta}_n),
\label{eq:accelerated_forward_model}
\end{equation}
where $a_0\equiv1$ selects the mean beam, $T_{21}$ is the global 21-cm signal, and $n(\nu,t;\boldsymbol{\theta}_n)$ denotes a realisation of the adopted noise model, parameterised by $\boldsymbol{\theta}_{\mathrm{Noise}}\equiv\boldsymbol{\theta}_n$ \citep{Scheutwinkel_2022a}.

End-to-end uncertainty propagation is thereby reduced to a compact neural-network evaluation and tensor contractions, yielding a low-memory, differentiable, and JIT-compilable forward model that is parallelisable across frequency, time, and simulations. On a single GPU, this implementation reduces the $\sim2.8$-hour full-fidelity runtime to $0.140\pm0.002~\mathrm{ms}$, a factor of $7.2\times10^{7}$.\footnote{\label{fn:h100_benchmark}Timings report the mean $\pm$ standard deviation across 500 warm evaluations on a single NVIDIA H100 GPU with 96~GB of HBM3. Costs assume \pounds2.94 per H100-hour based on \href{https://aws.amazon.com/ec2/pricing/on-demand/}{AWS pricing}.} Vectorising batches of 512 fully exploits GPU parallelism, reaching 384,939 simulations per second at a marginal cost of $2.1\times10^{-9}$ pounds per simulation, a $7.6\times10^{9}$-fold reduction.

\section{Benchmark Inference}
\label{sec:experiments}

Computational acceleration is only meaningful if the hybrid approximations adequately represent the relevant degeneracies, enabling accurate recovery and calibrated posterior uncertainty. At this dynamic range, residual surrogate errors can exceed the target signal, while inadequate physical constraints on nuisance components can leave it unidentifiable. We therefore benchmark end-to-end recovery using vectorised NSS \citep{yallup2025nested,cabezas2024blackjax}, jointly testing the fidelity of the accelerated simulator against full-fidelity observations and the sampler's exploration of the resulting posterior.\footnote{The synthetic benchmark datasets, NSS chains, and inference products are available in \href{https://zenodo.org/records/22046858}{Zenodo record 22046858}.} The benchmark comprises 100 unseen signal profiles spanning different levels of detectability and degeneracy with foreground and instrumental components. Each is injected into an observation generated using a full-fidelity sky realisation, a held-out CEM beam simulation, and a fixed three-hour observing window. The full formalism used to construct these datasets is detailed in Appendix~\ref{app:data_generation}.

The complete forward model considered here contains 137 uncertain parameters: 100 beam coefficients, $\boldsymbol{\theta}_{\mathrm{Beam}}$, 30 foreground spectral indices, $\boldsymbol{\theta}_{\mathrm{FG}}$, six astrophysical parameters, $\boldsymbol{\theta}_{21}$, and one Gaussian noise amplitude, $\boldsymbol{\theta}_{\mathrm{Noise}}$. By design, the beam model is linear in $\boldsymbol{a}$ (Equation~\ref{eq:beam_surrogate}). Assuming the Gaussian prior $\boldsymbol{a}\sim\mathcal{N}(\boldsymbol{0},\mathbf{\Sigma}_a)$, all 100 coefficients can thus be marginalised analytically, reducing the directly sampled space to 37 dimensions while retaining their contribution to the likelihood covariance:
\begin{equation}
\mathcal{L}_{\mathrm{marg}}(\mathbf{d}\mid\boldsymbol{\phi})
=
\mathcal{N}\!\left[
\mathbf{d}
\,\middle|\,
\boldsymbol{\mu}(\boldsymbol{\phi}),
\mathbf{C}_{n}(\theta_n)
+
\mathbf{H}(\boldsymbol{\phi})
\mathbf{\Sigma}_a
\mathbf{H}^{\mathrm{T}}(\boldsymbol{\phi})
\right],
\label{eq:marginalised_likelihood}
\end{equation}
where $\mathbf{d}$ is the observed data vector, $\mathbf{C}_{n}$ the noise covariance, $\boldsymbol{\phi}=(\boldsymbol{\theta}_{\mathrm{FG}},\boldsymbol{\theta}_{21},\boldsymbol{\theta}_{\mathrm{Noise}})$, $\boldsymbol{\mu}$ is the mean-beam forward model, and $\mathbf{H}$ maps the beam coefficients into the observed spectrum (derived in Appendix~\ref{app:beam_marginalisation}). Using this covariance form, the complete marginalised likelihood takes $0.291\pm0.003~\mathrm{ms}$ per evaluation and reaches 30,696 evaluations per second in batches of 2,048.\footnotemark[\getrefnumber{fn:h100_benchmark}]

\subsection{Signal recovery}
\label{sec:signal_recovery}

We evaluate recovery in signal space, since the intrinsic degeneracies and multimodality of the astrophysical parameterisation broaden the marginal parameter posteriors (Appendix~\ref{app:parameter_posteriors}) and would otherwise be conflated with inference performance. Signal-profile credible-band coverage evaluates posterior calibration, while posterior-mean RMSE independently quantifies the accuracy of the recovered signal, with both diagnostics defined in Appendix~\ref{app:recovery_metrics}.

Figure~\ref{fig:radio_galaxy_signal_recovery} shows representative profile recoveries. Across the complete benchmark, 94 of 100 injected profiles lie entirely within their 95\% whole-profile credible bands, consistent with nominal coverage, and 75 posterior-mean profiles have an RMSE below the assumed $25~\mathrm{mK}$ instrumental-noise level. Residual departures between the posterior mean and injected profile are therefore consistent with calibrated uncertainty, with the credible-band widths reflecting genuine degeneracies between the signal and the foreground and beam components. Taken together, these results show that the computational gains of the hybrid framework are achieved without compromising signal recovery or posterior calibration, even when no component of the observational chain is treated as known. The \href{https://github.com/JacobTutt/global-21cm-inference-benchmark}{open-source benchmark} therefore provides full-fidelity test observations, the precomputed response operator, and ready-to-run implementations of the accelerated simulator and its full and analytically marginalised likelihoods, forming a 100-realisation benchmark suite for alternative inference methods. \href{https://zenodo.org/records/22046858}{Reference NSS posteriors and evidence estimates} are released for all 100 realisations.

\begin{figure}[!t]
    \centering
    \includegraphics[width=\textwidth]{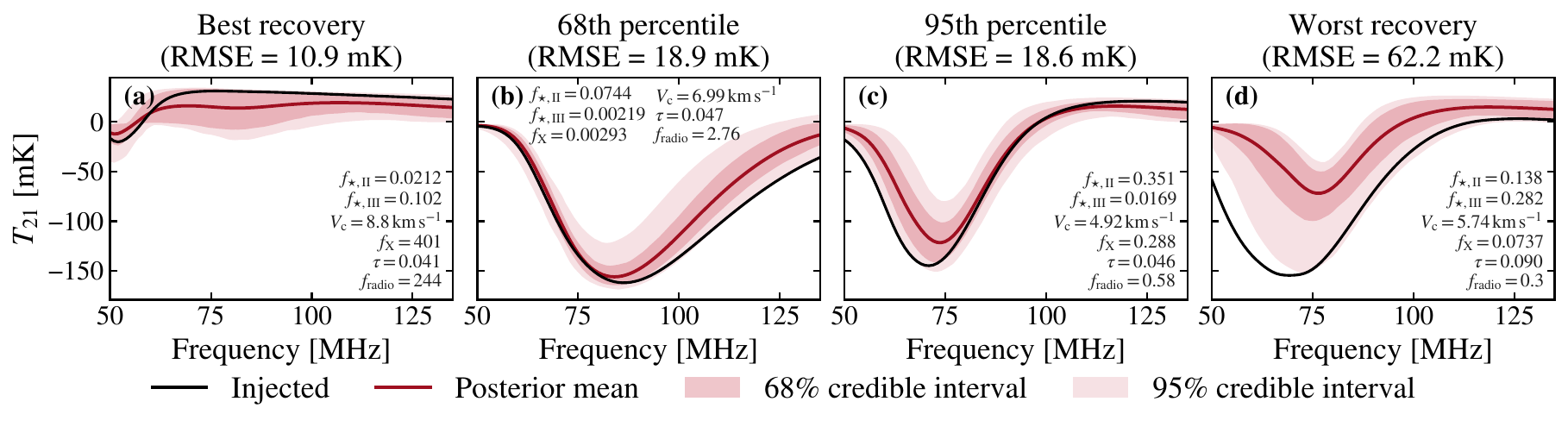}
    \caption{\textbf{Signal-profile recovery.} Representative cases ordered by minimum whole-profile credible level. Lines denote the injection and posterior mean, shading the pointwise 68\% and 95\% intervals, and titles the RMSE.}
    \label{fig:radio_galaxy_signal_recovery}
\end{figure}

\section{Limitations and Future Work}
\label{sec:limitations}

To our knowledge, this is the first work to jointly propagate the dominant uncertainties in global 21-cm inference in a physically motivated Bayesian forward-modelling framework, but a complete digital twin must also incorporate secondary observational effects \citep{Mittal_2024,Robins_2026,Shen_2021,Shen_2022,Pattison_2023,pattison_2025a,pattison_2025b}, receiver-calibration uncertainty \citep{Roque_2025_receiver,Kirkham_2025}, and radiometer noise beyond the homoskedastic Gaussian model adopted here. Existing formalisms support these extensions: additional physical components can be integrated modularly, while heteroskedastic noise is accommodated through a diagonal $\mathbf{C}_n$ set by the radiometer equation. These will nevertheless increase dimensionality and complicate the posterior geometry. This exacerbates the remaining inference bottleneck: NSS scales theoretically as $\mathcal{O}(D^3)$ \citep{polychord}, while the present 37-dimensional analyses already require 5~h~34~min to 7~h~16~min per realisation,\footnote{Totalling 594.4 GPU-hours across the 100 runs, each on one NVIDIA H100 GPU (96~GB HBM3).} making scalable posterior exploration and evidence-based model comparison a key challenge \citep{Sims_2025}.

While standard benchmark suites such as \texttt{sbibm} support controlled and reproducible comparison of inference methods \citep{Lueckmann_2021}, scientific applications are increasing in dimensionality and structural complexity. Examples include the transition from compact gravitational-wave source models to stochastic backgrounds in astrophysics \citep{Alvey_2024_SGWB}, and from point-neuron models to high-dimensional inference over detailed biophysical neurons and circuits in neuroscience \citep{Goncalves_2020,Deistler_2025_Jaxley,Tanoh_2025}. Designed for direct integration and rapid, low-cost evaluation, the framework supports benchmarking next-generation sampling algorithms \citep{yallup2026nestedsamplingslicewithingibbsefficient,yallup_2025_gw}, simulation-based inference \citep{Cranmer_2020,Deistler_2025_SBI_Guide,Glockler_2022,Lyu_2026_DynamicSBI,Alvey_2026,leeney_2026}, and variational inference \citep{Cai_2024_EigenVI,Gunapati_2022}, testing computational efficiency, posterior calibration, mode coverage, and evidence accuracy on a scientifically timely, high-dimensional inverse problem at the frontier of early-Universe cosmology.

\newpage
\begin{ack}
The authors acknowledge the use of resources provided by the Isambard-AI National AI Research Resource (AIRR). Isambard-AI \citep{Isamabrd_2024} is operated by the University of Bristol and is funded by the UK Government's Department for Science, Innovation and Technology (DSIT) via UK Research and Innovation; and the Science and Technology Facilities Council [ST/AIRR/I-A-I/1023]. This work used the DiRAC Data Intensive service (CSD3, project number APP52749) at the University of Cambridge, managed by the University of Cambridge University Information Services on behalf of the STFC DiRAC HPC Facility (\url{www.dirac.ac.uk}). The DiRAC component of CSD3 at Cambridge was funded by BEIS, UKRI and STFC capital funding and STFC operations grants. DiRAC is part of the UKRI Digital Research Infrastructure. JT is supported by the Harding Distinguished Postgraduate Scholars Programme (HDPSP) and the Science and Technology Facilities Council (STFC) DTP Studentship. HTJB and JA acknowledge support from the Kavli Institute for Cosmology Cambridge and the Kavli Foundation. HTJB also acknowledges support from Imperial College London and Schmidt Sciences through an Imperial College Research Fellowship. JMC and DJA were supported by the Science and Technology Facilities Council. We would also like to thank the Kavli Foundation for their support of REACH. EdLA acknowledges the support of STFC via an Ernest Rutherford Fellowship. This work was supported by UK Research and Innovation (UKRI) through the Horizon Europe Guarantee under grant EP/Y02916X/1 (REACH 21).
\end{ack}

\bibliographystyle{unsrtnat}
\bibliography{references}

\begin{thebibliography}{86}
\providecommand{\natexlab}[1]{#1}
\providecommand{\url}[1]{\texttt{#1}}
\expandafter\ifx\csname urlstyle\endcsname\relax
  \providecommand{\doi}[1]{doi: #1}\else
  \providecommand{\doi}{doi: \begingroup \urlstyle{rm}\Url}\fi

\bibitem[Furlanetto et~al.(2006)Furlanetto, Peng~Oh, and
  Briggs]{Furlanetto_2006}
Steven~R. Furlanetto, S.~Peng~Oh, and Frank~H. Briggs.
\newblock Cosmology at low frequencies: The 21cm transition and the
  high-redshift universe.
\newblock \emph{Physics Reports}, 433\penalty0 (4–6):\penalty0 181–301,
  October 2006.
\newblock ISSN 0370-1573.
\newblock \doi{10.1016/j.physrep.2006.08.002}.
\newblock URL \url{http://dx.doi.org/10.1016/j.physrep.2006.08.002}.

\bibitem[Pritchard and Loeb(2012)]{Pritchard_2012}
Jonathan~R Pritchard and Abraham Loeb.
\newblock 21 cm cosmology in the 21st century.
\newblock \emph{Reports on Progress in Physics}, 75\penalty0 (8):\penalty0
  086901, July 2012.
\newblock ISSN 1361-6633.
\newblock \doi{10.1088/0034-4885/75/8/086901}.
\newblock URL \url{http://dx.doi.org/10.1088/0034-4885/75/8/086901}.

\bibitem[Bowman et~al.(2018)Bowman, Rogers, Monsalve, Mozdzen, and
  Mahesh]{Bowman_2018}
Judd~D. Bowman, Alan E.~E. Rogers, Raul~A. Monsalve, Thomas~J. Mozdzen, and
  Nivedita Mahesh.
\newblock An absorption profile centred at 78 megahertz in the sky-averaged
  spectrum.
\newblock \emph{Nature}, 555\penalty0 (7694):\penalty0 67–70, March 2018.
\newblock ISSN 1476-4687.
\newblock \doi{10.1038/nature25792}.
\newblock URL \url{http://dx.doi.org/10.1038/nature25792}.

\bibitem[Philip et~al.(2019)Philip, Abdurashidova, Chiang, Ghazi, Gumba,
  Heilgendorff, Jáuregui~Garcia, Malepe, Nunhokee, Peterson, Sievers, Simes,
  and Spann]{Philip_2018}
Liju Philip, Z.~Abdurashidova, H.~Chiang, N.~Ghazi, A.~Gumba, H.~Heilgendorff,
  Jose Jáuregui~Garcia, Kagiso Malepe, C.~Nunhokee, Jaron Peterson,
  J.~Sievers, V.~Simes, and R.~Spann.
\newblock Probing radio intensity at {High-Z} from {Marion}: 2017 instrument.
\newblock \emph{Journal of Astronomical Instrumentation}, 08\penalty0
  (2):\penalty0 1950004, 2019.
\newblock \doi{10.1142/S2251171719500041}.

\bibitem[{Singh} et~al.(2018){Singh}, {Subrahmanyan}, {Udaya Shankar},
  {Sathyanarayana Rao}, {Fialkov}, {Cohen}, {Barkana}, {Girish}, {Raghunathan},
  {Somashekar}, and {Srivani}]{Singh_2018}
Saurabh {Singh}, Ravi {Subrahmanyan}, N.~{Udaya Shankar}, Mayuri
  {Sathyanarayana Rao}, Anastasia {Fialkov}, Aviad {Cohen}, Rennan {Barkana},
  B.~S. {Girish}, A.~{Raghunathan}, R.~{Somashekar}, and K.~S. {Srivani}.
\newblock {SARAS 2 Constraints on Global 21 cm Signals from the Epoch of
  Reionization}.
\newblock \emph{\apj}, 858\penalty0 (1):\penalty0 54, May 2018.
\newblock \doi{10.3847/1538-4357/aabae1}.

\bibitem[Monsalve et~al.(2024)Monsalve, Bye, Sievers, Bidula, Bustos, Chiang,
  Guo, Hendricksen, McGee, Mena, Prabhakar, Restrepo, and
  Thyagarajan]{Monsalve_2024}
Raul~A. Monsalve, Christian~H. Bye, Jonathan~L. Sievers, Vadym Bidula, Ricardo
  Bustos, H.~Cynthia Chiang, Xinze Guo, Ian Hendricksen, Francis McGee,
  F.~Patricio Mena, Garima Prabhakar, Oscar Restrepo, and Nithyanandan
  Thyagarajan.
\newblock Simulating the detection of the global 21 cm signal with {MIST} for
  different models of the soil and beam directivity.
\newblock \emph{The Astrophysical Journal}, 961\penalty0 (1):\penalty0 56,
  January 2024.
\newblock ISSN 1538-4357.
\newblock \doi{10.3847/1538-4357/ad0f1b}.
\newblock URL \url{http://dx.doi.org/10.3847/1538-4357/ad0f1b}.

\bibitem[Bull et~al.(2025)Bull, El-Makadema, Garsden, Edgley, Roddis, Chluba,
  Conselice, Dutta, Glasscock, Nasirudin, Norris, Wilensky, Ye, and
  Zhang]{Bull_2025}
Philip Bull, Ahmed El-Makadema, Hugh Garsden, John Edgley, Neil Roddis, Jens
  Chluba, Christopher~J Conselice, Sohini Dutta, Katrine~A Glasscock,
  Ainulnabilah Nasirudin, Jordan Norris, Michael~J Wilensky, Isabelle Ye, and
  Zheng Zhang.
\newblock {RHINO}: a large horn antenna for detecting the 21 cm global
  signal.
\newblock \emph{RAS Techniques and Instruments}, 4:\penalty0 rzaf046, 2025.
\newblock ISSN 2752-8200.
\newblock \doi{10.1093/rasti/rzaf046}.
\newblock URL \url{http://dx.doi.org/10.1093/rasti/rzaf046}.

\bibitem[de~Lera~Acedo et~al.(2022)de~Lera~Acedo, de~Villiers, Razavi-Ghods,
  Handley, Fialkov, Magro, Anstey, Bevins, Chiello, Cumner, Josaitis, Roque,
  Sims, Scheutwinkel, Alexander, Bernardi, Carey, Cavillot, Croukamp, Ely,
  Gessey-Jones, Gueuning, Hills, Kulkarni, Maiolino, Meerburg, Mittal,
  Pritchard, Puchwein, Saxena, Shen, Smirnov, Spinelli, and
  Zarb-Adami]{Acedo_2022}
E.~de~Lera~Acedo, D.~I.~L. de~Villiers, N.~Razavi-Ghods, W.~Handley,
  A.~Fialkov, A.~Magro, D.~Anstey, H.~T.~J. Bevins, R.~Chiello, J.~Cumner,
  A.~T. Josaitis, I.~L.~V. Roque, P.~H. Sims, K.~H. Scheutwinkel, P.~Alexander,
  G.~Bernardi, S.~Carey, J.~Cavillot, W.~Croukamp, J.~A. Ely, T.~Gessey-Jones,
  Q.~Gueuning, R.~Hills, G.~Kulkarni, R.~Maiolino, P.~D. Meerburg, S.~Mittal,
  J.~R. Pritchard, E.~Puchwein, A.~Saxena, E.~Shen, O.~Smirnov, M.~Spinelli,
  and K.~Zarb-Adami.
\newblock The {REACH} radiometer for detecting the 21-cm hydrogen signal from
  redshift z ≈ 7.5–28.
\newblock \emph{Nature Astronomy}, 6\penalty0 (8):\penalty0 984–998, July
  2022.
\newblock ISSN 2397-3366.
\newblock \doi{10.1038/s41550-022-01709-9}.
\newblock URL \url{http://dx.doi.org/10.1038/s41550-022-01709-9}.

\bibitem[{Harker} et~al.(2012){Harker}, {Pritchard}, {Burns}, and
  {Bowman}]{Harker_2012}
Geraint J.~A. {Harker}, Jonathan~R. {Pritchard}, Jack~O. {Burns}, and Judd~D.
  {Bowman}.
\newblock {An MCMC approach to extracting the global 21-cm signal during the
  cosmic dawn from sky-averaged radio observations}.
\newblock \emph{\mnras}, 419\penalty0 (2):\penalty0 1070--1084, January 2012.
\newblock \doi{10.1111/j.1365-2966.2011.19766.x}.

\bibitem[Liu et~al.(2013)Liu, Pritchard, Tegmark, and Loeb]{Liu_2013}
Adrian Liu, Jonathan~R. Pritchard, Max Tegmark, and Abraham Loeb.
\newblock Global 21 cm signal experiments: A designer's guide.
\newblock \emph{Physical Review D}, 87\penalty0 (4):\penalty0 043002, February
  2013.
\newblock ISSN 1550-2368.
\newblock \doi{10.1103/physrevd.87.043002}.
\newblock URL \url{http://dx.doi.org/10.1103/PhysRevD.87.043002}.

\bibitem[Switzer and Liu(2014)]{Switzer_2014}
Eric~R. Switzer and Adrian Liu.
\newblock Erasing the variable: Empirical foreground discovery for global 21 cm
  spectrum experiments.
\newblock \emph{The Astrophysical Journal}, 793\penalty0 (2):\penalty0 102,
  September 2014.
\newblock ISSN 1538-4357.
\newblock \doi{10.1088/0004-637x/793/2/102}.
\newblock URL \url{http://dx.doi.org/10.1088/0004-637X/793/2/102}.

\bibitem[Monsalve et~al.(2017)Monsalve, Rogers, Bowman, and
  Mozdzen]{Monsalve_2017}
Raul~A. Monsalve, Alan E.~E. Rogers, Judd~D. Bowman, and Thomas~J. Mozdzen.
\newblock Results from edges high-band. i. constraints on phenomenological
  models for the global 21 cm signal.
\newblock \emph{The Astrophysical Journal}, 847\penalty0 (1):\penalty0 64,
  September 2017.
\newblock \doi{10.3847/1538-4357/aa88d1}.
\newblock URL \url{https://doi.org/10.3847/1538-4357/aa88d1}.

\bibitem[Tauscher et~al.(2018{\natexlab{a}})Tauscher, Rapetti, Burns, and
  Switzer]{Tauscher_2018a}
Keith Tauscher, David Rapetti, Jack~O. Burns, and Eric Switzer.
\newblock Global 21 cm signal extraction from foreground and instrumental
  effects. {I}. pattern recognition framework for separation using training
  sets.
\newblock \emph{The Astrophysical Journal}, 853\penalty0 (2):\penalty0 187,
  February 2018{\natexlab{a}}.
\newblock ISSN 1538-4357.
\newblock \doi{10.3847/1538-4357/aaa41f}.
\newblock URL \url{http://dx.doi.org/10.3847/1538-4357/aaa41f}.

\bibitem[Tauscher et~al.(2018{\natexlab{b}})Tauscher, Rapetti, and
  Burns]{Tauscher_2018b}
K.~Tauscher, D.~Rapetti, and J.O. Burns.
\newblock A new goodness-of-fit statistic and its application to 21-cm
  cosmology.
\newblock \emph{Journal of Cosmology and Astroparticle Physics}, 2018\penalty0
  (12):\penalty0 015--015, December 2018{\natexlab{b}}.
\newblock ISSN 1475-7516.
\newblock \doi{10.1088/1475-7516/2018/12/015}.
\newblock URL \url{http://dx.doi.org/10.1088/1475-7516/2018/12/015}.

\bibitem[Hills et~al.(2018)Hills, Kulkarni, Meerburg, and Puchwein]{Hills_2018}
Richard Hills, Girish Kulkarni, P.~Daniel Meerburg, and Ewald Puchwein.
\newblock Concerns about modelling of the edges data.
\newblock \emph{Nature}, 564\penalty0 (7736):\penalty0 E32–E34, December
  2018.
\newblock ISSN 1476-4687.
\newblock \doi{10.1038/s41586-018-0796-5}.
\newblock URL \url{http://dx.doi.org/10.1038/s41586-018-0796-5}.

\bibitem[Singh and Subrahmanyan(2019)]{Singh_2019}
Saurabh Singh and Ravi Subrahmanyan.
\newblock The redshifted 21 cm signal in the {EDGES} low-band spectrum.
\newblock \emph{The Astrophysical Journal}, 880\penalty0 (1):\penalty0 26, July
  2019.
\newblock ISSN 1538-4357.
\newblock \doi{10.3847/1538-4357/ab2879}.
\newblock URL \url{http://dx.doi.org/10.3847/1538-4357/ab2879}.

\bibitem[{Sims} and {Pober}(2020)]{Sims_2020}
Peter~H. {Sims} and Jonathan~C. {Pober}.
\newblock {Testing for calibration systematics in the EDGES low-band data using
  Bayesian model selection}.
\newblock \emph{\mnras}, 492\penalty0 (1):\penalty0 22--38, February 2020.
\newblock \doi{10.1093/mnras/stz3388}.

\bibitem[Rapetti et~al.(2020)Rapetti, Tauscher, Mirocha, and
  Burns]{Rapetti_2020}
David Rapetti, Keith Tauscher, Jordan Mirocha, and Jack~O. Burns.
\newblock Global 21 cm signal extraction from foreground and instrumental
  effects. {II}. efficient and self-consistent technique for constraining
  nonlinear signal models.
\newblock \emph{The Astrophysical Journal}, 897\penalty0 (2):\penalty0 174,
  July 2020.
\newblock ISSN 1538-4357.
\newblock \doi{10.3847/1538-4357/ab9b29}.
\newblock URL \url{http://dx.doi.org/10.3847/1538-4357/ab9b29}.

\bibitem[Tauscher et~al.(2020{\natexlab{a}})Tauscher, Rapetti, and
  Burns]{Tauscher_2020a}
Keith Tauscher, David Rapetti, and Jack~O. Burns.
\newblock Global 21 cm signal extraction from foreground and instrumental
  effects. {III}. utilizing drift-scan time dependence and full stokes
  measurements.
\newblock \emph{The Astrophysical Journal}, 897\penalty0 (2):\penalty0 175,
  July 2020{\natexlab{a}}.
\newblock ISSN 1538-4357.
\newblock \doi{10.3847/1538-4357/ab9b2a}.
\newblock URL \url{http://dx.doi.org/10.3847/1538-4357/ab9b2a}.

\bibitem[Tauscher et~al.(2020{\natexlab{b}})Tauscher, Rapetti, and
  Burns]{Tauscher_2020b}
Keith Tauscher, David Rapetti, and Jack~O. Burns.
\newblock Formulating and critically examining the assumptions of global 21 cm
  signal analyses: How to avoid the false troughs that can appear in
  single-spectrum fits.
\newblock \emph{The Astrophysical Journal}, 897\penalty0 (2):\penalty0 132,
  July 2020{\natexlab{b}}.
\newblock ISSN 1538-4357.
\newblock \doi{10.3847/1538-4357/ab9a3f}.
\newblock URL \url{http://dx.doi.org/10.3847/1538-4357/ab9a3f}.

\bibitem[Tauscher et~al.(2021)Tauscher, Rapetti, Nhan, Handy, Bassett, Hibbard,
  Bordenave, Bradley, and Burns]{Tauscher_2021}
Keith Tauscher, David Rapetti, Bang~D. Nhan, Alec Handy, Neil Bassett, Joshua
  Hibbard, David Bordenave, Richard~F. Bradley, and Jack~O. Burns.
\newblock Global 21 cm signal extraction from foreground and instrumental
  effects. {IV}. accounting for realistic instrument uncertainties and their
  overlap with foreground and signal models.
\newblock \emph{The Astrophysical Journal}, 915\penalty0 (1):\penalty0 66, July
  2021.
\newblock ISSN 1538-4357.
\newblock \doi{10.3847/1538-4357/ac00af}.
\newblock URL \url{http://dx.doi.org/10.3847/1538-4357/ac00af}.

\bibitem[Bevins et~al.(2021{\natexlab{a}})Bevins, Handley, Fialkov,
  de Lera Acedo, Greenhill, and Price]{Bevins_2021}
H~T~J Bevins, W~J Handley, A~Fialkov, E~de Lera Acedo, L~J Greenhill, and D~C
  Price.
\newblock <scp>maxsmooth</scp>: rapid maximally smooth function fitting with
  applications in global 21-cm cosmology.
\newblock \emph{Monthly Notices of the Royal Astronomical Society},
  502\penalty0 (3):\penalty0 4405–4425, January 2021{\natexlab{a}}.
\newblock ISSN 1365-2966.
\newblock \doi{10.1093/mnras/stab152}.
\newblock URL \url{http://dx.doi.org/10.1093/mnras/stab152}.

\bibitem[Anstey et~al.(2023)Anstey, de Lera Acedo, and Handley]{Anstey_2022}
Dominic Anstey, Eloy de Lera Acedo, and Will Handley.
\newblock Use of time dependent data in bayesian global 21-cm foreground and
  signal modelling.
\newblock \emph{Monthly Notices of the Royal Astronomical Society},
  520\penalty0 (1):\penalty0 850--865, 01 2023.
\newblock ISSN 0035-8711.
\newblock \doi{10.1093/mnras/stad156}.
\newblock URL \url{https://doi.org/10.1093/mnras/stad156}.

\bibitem[Murray et~al.(2022)Murray, Bowman, Sims, Mahesh, Rogers, Monsalve,
  Samson, and Vydula]{Murray_2022}
Steven~G. Murray, Judd~D. Bowman, Peter~H. Sims, Nivedita Mahesh, Alan E.~E.
  Rogers, Raul~A. Monsalve, Titu Samson, and Akshatha~Konakondula Vydula.
\newblock A bayesian calibration framework for {EDGES}.
\newblock \emph{Monthly Notices of the Royal Astronomical Society},
  517\penalty0 (2):\penalty0 2264--2284, September 2022.
\newblock ISSN 1365-2966.
\newblock \doi{10.1093/mnras/stac2600}.
\newblock URL \url{http://dx.doi.org/10.1093/mnras/stac2600}.

\bibitem[Saxena et~al.(2023)Saxena, Meerburg, de Lera Acedo, Handley, and
  Koopmans]{Saxena_2023}
Anchal Saxena, P~Daniel Meerburg, Eloy de Lera Acedo, Will Handley, and Léon
  V~E Koopmans.
\newblock Sky-averaged 21-cm signal extraction using multiple antennas with an
  {SVD} framework: the {REACH} case.
\newblock \emph{Monthly Notices of the Royal Astronomical Society},
  522\penalty0 (1):\penalty0 1022–1032, April 2023.
\newblock ISSN 1365-2966.
\newblock \doi{10.1093/mnras/stad1047}.
\newblock URL \url{http://dx.doi.org/10.1093/mnras/stad1047}.

\bibitem[Hibbard et~al.(2023)Hibbard, Rapetti, Burns, Mahesh, and
  Bassett]{Hibbard_2023}
Joshua~J. Hibbard, David Rapetti, Jack~O. Burns, Nivedita Mahesh, and Neil
  Bassett.
\newblock Fitting and comparing galactic foreground models for unbiased 21 cm
  cosmology.
\newblock \emph{The Astrophysical Journal}, 959\penalty0 (2):\penalty0 103,
  December 2023.
\newblock ISSN 1538-4357.
\newblock \doi{10.3847/1538-4357/ad06b3}.
\newblock URL \url{http://dx.doi.org/10.3847/1538-4357/ad06b3}.

\bibitem[Saxena et~al.(2024)Saxena, Meerburg, Weniger, Acedo, and
  Handley]{saxena_2024}
Anchal Saxena, P~Daniel Meerburg, Christoph Weniger, Eloy de~Lera Acedo, and
  Will Handley.
\newblock Simulation-based inference of the sky-averaged 21-cm signal from
  {CD-EoR} with {REACH}.
\newblock \emph{RAS Techniques and Instruments}, 3\penalty0 (1):\penalty0
  724–736, Jan 2024.
\newblock ISSN 2752-8200.
\newblock \doi{10.1093/rasti/rzae047}.
\newblock URL \url{http://dx.doi.org/10.1093/rasti/rzae047}.

\bibitem[Sims et~al.(2025{\natexlab{a}})Sims, Bowman, Murray, Barrett,
  Cappallo, Lonsdale, Mahesh, Monsalve, Rogers, Samson, and Vydula]{Sims_2025}
Peter~H Sims, Judd~D Bowman, Steven~G Murray, John~P Barrett, Rigel~C Cappallo,
  Colin~J Lonsdale, Nivedita Mahesh, Raul~A Monsalve, Alan E~E Rogers, Titu
  Samson, and Akshatha~K Vydula.
\newblock A general bayesian model-validation framework based on null-test
  evidence ratios, with an example application to global 21-cm cosmology.
\newblock \emph{Monthly Notices of the Royal Astronomical Society},
  541\penalty0 (3):\penalty0 2262–2281, July 2025{\natexlab{a}}.
\newblock ISSN 1365-2966.
\newblock \doi{10.1093/mnras/staf1109}.
\newblock URL \url{http://dx.doi.org/10.1093/mnras/staf1109}.

\bibitem[Sims et~al.(2025{\natexlab{b}})Sims, Bowman, Murray, Barrett,
  Cappallo, Lonsdale, Mahesh, Monsalve, Rogers, Samson, and Vydula]{sims_2023}
Peter~H Sims, Judd~D Bowman, Steven~G Murray, John~P Barrett, Rigel~C Cappallo,
  Colin~J Lonsdale, Nivedita Mahesh, Raul~A Monsalve, Alan E~E Rogers, Titu
  Samson, and Akshatha~K Vydula.
\newblock A bayesian approach to modelling spectrometer data chromaticity
  corrected using beam factors – {II}. model priors and posterior odds.
\newblock \emph{Monthly Notices of the Royal Astronomical Society},
  544\penalty0 (2):\penalty0 2340--2364, 12 2025{\natexlab{b}}.
\newblock ISSN 0035-8711.
\newblock \doi{10.1093/mnras/staf1767}.
\newblock URL \url{https://doi.org/10.1093/mnras/staf1767}.

\bibitem[Tutt et~al.(2026{\natexlab{a}})Tutt, Sims, Pattison, Anstey, Leeney,
  and de Lera Acedo]{Tutt_2026a}
Jacob~L Tutt, Peter~H Sims, Joe H~N Pattison, Dominic~J Anstey, Samuel A~K
  Leeney, and Eloy de Lera Acedo.
\newblock Optimizing foreground modelling for global 21-cm cosmology with
  {GPU}-accelerated nested sampling.
\newblock \emph{Monthly Notices of the Royal Astronomical Society},
  550\penalty0 (1):\penalty0 stag1101, June 2026{\natexlab{a}}.
\newblock ISSN 1365-2966.
\newblock \doi{10.1093/mnras/stag1101}.
\newblock URL \url{http://dx.doi.org/10.1093/mnras/stag1101}.

\bibitem[{Fialkov} et~al.(2014){Fialkov}, {Barkana}, and
  {Visbal}]{Fialkov_2014}
Anastasia {Fialkov}, Rennan {Barkana}, and Eli {Visbal}.
\newblock {The observable signature of late heating of the Universe during
  cosmic reionization}.
\newblock \emph{\nat}, 506\penalty0 (7487):\penalty0 197--199, February 2014.
\newblock \doi{10.1038/nature12999}.

\bibitem[{Shaver} et~al.(1999){Shaver}, {Windhorst}, {Madau}, and {de
  Bruyn}]{Shaver_1999}
P.~A. {Shaver}, R.~A. {Windhorst}, P.~{Madau}, and A.~G. {de Bruyn}.
\newblock {Can the reionization epoch be detected as a global signature in the
  cosmic background?}
\newblock \emph{\aap}, 345:\penalty0 380--390, May 1999.
\newblock \doi{10.48550/arXiv.astro-ph/9901320}.

\bibitem[Anstey et~al.(2021)Anstey, de Lera Acedo, and Handley]{Anstey_2021}
Dominic Anstey, Eloy de Lera Acedo, and Will Handley.
\newblock A general bayesian framework for foreground modelling and
  chromaticity correction for global 21 cm experiments.
\newblock \emph{Monthly Notices of the Royal Astronomical Society},
  506\penalty0 (2):\penalty0 2041--2058, 06 2021.
\newblock ISSN 0035-8711.
\newblock \doi{10.1093/mnras/stab1765}.
\newblock URL \url{https://doi.org/10.1093/mnras/stab1765}.

\bibitem[Hibbard et~al.(2020)Hibbard, Tauscher, Rapetti, and
  Burns]{Hibbard_2020}
Joshua~J. Hibbard, Keith Tauscher, David Rapetti, and Jack~O. Burns.
\newblock Modeling the galactic foreground and beam chromaticity for global 21
  cm cosmology.
\newblock \emph{The Astrophysical Journal}, 905\penalty0 (2):\penalty0 113,
  December 2020.
\newblock ISSN 1538-4357.
\newblock \doi{10.3847/1538-4357/abc3c5}.
\newblock URL \url{http://dx.doi.org/10.3847/1538-4357/abc3c5}.

\bibitem[Bassett et~al.(2021)Bassett, Rapetti, Tauscher, Burns, and
  Hibbard]{Bassett_2021a}
Neil Bassett, David Rapetti, Keith Tauscher, Jack~O. Burns, and Joshua~J.
  Hibbard.
\newblock Ensuring robustness in training-set-based global 21 cm cosmology
  analysis.
\newblock \emph{The Astrophysical Journal}, 908\penalty0 (2):\penalty0 189,
  February 2021.
\newblock ISSN 1538-4357.
\newblock \doi{10.3847/1538-4357/abdb29}.
\newblock URL \url{http://dx.doi.org/10.3847/1538-4357/abdb29}.

\bibitem[Cumner et~al.(2024)Cumner, Pieterse, de~Villiers, and
  de~Lera~Acedo]{Cumner_2024}
John Cumner, Carla Pieterse, Dirk de~Villiers, and Eloy de~Lera~Acedo.
\newblock The effects of the antenna power pattern uncertainty within a global
  21 cm experiment.
\newblock \emph{Monthly Notices of the Royal Astronomical Society},
  531\penalty0 (4):\penalty0 4734--4745, June 2024.
\newblock \doi{10.1093/mnras/stae1475}.
\newblock URL \url{https://doi.org/10.1093/mnras/stae1475}.

\bibitem[Pattison et~al.(2026{\natexlab{a}})Pattison, Cumner, Anstey, Pegwal,
  Croukamp, de~Villiers, and de~Lera~Acedo]{Pattison_2026}
Joe H.~N. Pattison, John~M. Cumner, Dominic~J. Anstey, Saurabh Pegwal, Wessel
  Croukamp, Dirk I.~L. de~Villiers, and Eloy de~Lera~Acedo.
\newblock Impact of antenna structure and orientation on forward-modelled
  global 21 cm signal recovery.
\newblock \emph{Monthly Notices of the Royal Astronomical Society},
  550\penalty0 (2):\penalty0 stag1216, 2026{\natexlab{a}}.
\newblock \doi{10.1093/mnras/stag1216}.
\newblock URL \url{https://doi.org/10.1093/mnras/stag1216}.

\bibitem[Cohen et~al.(2017)Cohen, Fialkov, Barkana, and Lotem]{Cohen_2017}
Aviad Cohen, Anastasia Fialkov, Rennan Barkana, and Matan Lotem.
\newblock Charting the parameter space of the global 21-cm signal.
\newblock \emph{Monthly Notices of the Royal Astronomical Society},
  472\penalty0 (2):\penalty0 1915--1931, 08 2017.
\newblock ISSN 0035-8711.
\newblock \doi{10.1093/mnras/stx2065}.
\newblock URL \url{https://doi.org/10.1093/mnras/stx2065}.

\bibitem[Harrington(1993)]{Harrington1993}
R.~F. Harrington.
\newblock \emph{Field Computation by Moment Methods}.
\newblock Wiley--IEEE Press, Hoboken, New Jersey, US, 1993.

\bibitem[Cranmer et~al.(2020)Cranmer, Brehmer, and Louppe]{Cranmer_2020}
Kyle Cranmer, Johann Brehmer, and Gilles Louppe.
\newblock The frontier of simulation-based inference.
\newblock \emph{Proceedings of the National Academy of Sciences}, 117\penalty0
  (48):\penalty0 30055--30062, 2020.
\newblock \doi{10.1073/pnas.1912789117}.
\newblock URL \url{https://doi.org/10.1073/pnas.1912789117}.

\bibitem[Yallup et~al.(2026)Yallup, Kroupa, and Handley]{yallup2025nested}
David Yallup, Namu Kroupa, and Will Handley.
\newblock Nested slice sampling: Vectorized nested sampling for
  {GPU}-accelerated inference.
\newblock \emph{Transactions on Machine Learning Research}, May 2026.
\newblock URL \url{https://openreview.net/forum?id=5mF2eRl3gt}.

\bibitem[Cabezas et~al.(2024)Cabezas, Corenflos, Lao, and
  Louf]{cabezas2024blackjax}
Alberto Cabezas, Adrien Corenflos, Junpeng Lao, and Rémi Louf.
\newblock Blackjax: Composable {B}ayesian inference in {JAX}, 2024.

\bibitem[Reis et~al.(2020)Reis, Fialkov, and Barkana]{Reis_2020_radio}
Itamar Reis, Anastasia Fialkov, and Rennan Barkana.
\newblock High-redshift radio galaxies: a potential new source of 21-cm
  fluctuations.
\newblock \emph{Monthly Notices of the Royal Astronomical Society},
  499\penalty0 (4):\penalty0 5993--6008, 10 2020.
\newblock \doi{10.1093/mnras/staa3091}.

\bibitem[Bevins et~al.(2021{\natexlab{b}})Bevins, Handley, Fialkov,
  de Lera Acedo, and Javid]{Bevins_2021_globalemu}
H~T~J Bevins, W~J Handley, A~Fialkov, E~de Lera Acedo, and K~Javid.
\newblock <scp>globalemu</scp>: a novel and robust approach for emulating the
  sky-averaged 21-cm signal from the cosmic dawn and epoch of reionization.
\newblock \emph{Monthly Notices of the Royal Astronomical Society},
  508\penalty0 (2):\penalty0 2923–2936, September 2021{\natexlab{b}}.
\newblock ISSN 1365-2966.
\newblock \doi{10.1093/mnras/stab2737}.
\newblock URL \url{http://dx.doi.org/10.1093/mnras/stab2737}.

\bibitem[Bevins et~al.(2025)Bevins, Gessey-Jones, and Handley]{Bevins_2025}
H~T~J Bevins, T~Gessey-Jones, and W~J Handley.
\newblock On the accuracy of posterior recovery with neural network emulators.
\newblock \emph{Monthly Notices of the Royal Astronomical Society},
  544\penalty0 (1):\penalty0 375–390, September 2025.
\newblock ISSN 1365-2966.
\newblock \doi{10.1093/mnras/staf1590}.
\newblock URL \url{http://dx.doi.org/10.1093/mnras/staf1590}.

\bibitem[{de Oliveira-Costa} et~al.(2008){de Oliveira-Costa}, {Tegmark},
  {Gaensler}, {Jonas}, {Landecker}, and {Reich}]{deOliveiraCosta2008}
Ang{\'e}lica {de Oliveira-Costa}, Max {Tegmark}, B.~M. {Gaensler}, Justin
  {Jonas}, T.~L. {Landecker}, and Patricia {Reich}.
\newblock {A model of diffuse Galactic radio emission from 10 MHz to 100 GHz}.
\newblock \emph{\mnras}, 388\penalty0 (1):\penalty0 247--260, July 2008.
\newblock \doi{10.1111/j.1365-2966.2008.13376.x}.

\bibitem[Pagano et~al.(2023)Pagano, Sims, Liu, Anstey, Handley, and
  de Lera Acedo]{Pagano_2023}
Michael Pagano, Peter Sims, Adrian Liu, Dominic Anstey, Will Handley, and Eloy
  de Lera Acedo.
\newblock A general bayesian framework to account for foreground map errors in
  global 21-cm experiments.
\newblock \emph{Monthly Notices of the Royal Astronomical Society},
  527\penalty0 (3):\penalty0 5649–5667, November 2023.
\newblock ISSN 1365-2966.
\newblock \doi{10.1093/mnras/stad3392}.
\newblock URL \url{http://dx.doi.org/10.1093/mnras/stad3392}.

\bibitem[Cavillot et~al.(2021)Cavillot, Craeye, de~Lera~Acedo, and
  Razavi-Ghods]{Cavillot_2021}
Jean Cavillot, Christophe Craeye, Eloy de~Lera~Acedo, and Nima Razavi-Ghods.
\newblock Efficient {MoM} simulation of {3D} metallic antenna connected to
  finite ground plane.
\newblock In \emph{2021 IEEE International Conference on Microwaves, Antennas,
  Communications and Electronic Systems (COMCAS)}, pages 213--216, 2021.
\newblock \doi{10.1109/COMCAS52219.2021.9629009}.
\newblock URL \url{https://doi.org/10.1109/COMCAS52219.2021.9629009}.

\bibitem[Hu et~al.(1999)Hu, Chew, Michielssen, and Zhao]{Hu1999}
B.~Hu, W.~C. Chew, E.~Michielssen, and J.~Zhao.
\newblock Fast inhomogeneous plane wave algorithm for the fast analysis of
  two-dimensional scattering problems.
\newblock \emph{Radio Science}, 34\penalty0 (4):\penalty0 759--772, 1999.
\newblock \doi{10.1029/1999RS900038}.

\bibitem[Alkhalifeh et~al.(2016)Alkhalifeh, Hislop, Ozdemir, and
  Craeye]{Alkhalifeh2016}
K.~Alkhalifeh, G.~Hislop, N.~A. Ozdemir, and C.~Craeye.
\newblock Efficient mom simulation of 3-d antennas in the vicinity of the
  ground.
\newblock \emph{IEEE Transactions on Antennas and Propagation}, 64\penalty0
  (12):\penalty0 5335--5344, 2016.
\newblock \doi{10.1109/TAP.2016.2618482}.

\bibitem[Cavillot et~al.(2020)Cavillot, Tihon, Mesa, de~Lera~Acedo, and
  Craeye]{Cavillot2020}
Jean Cavillot, Denis Tihon, Francisco Mesa, Eloy de~Lera~Acedo, and Christophe
  Craeye.
\newblock Efficient simulation of large irregular arrays on a finite ground
  plane.
\newblock \emph{IEEE Transactions on Antennas and Propagation}, 68:\penalty0
  2753--2764, 2020.
\newblock \doi{10.1109/TAP.2019.2955180}.

\bibitem[Cavillot et~al.(2024)Cavillot, Tihon, Gueuning, de~Lera~Acedo, and
  Craeye]{Cavillot2024}
Jean Cavillot, Denis Tihon, Quentin Gueuning, Eloy de~Lera~Acedo, and
  Christophe Craeye.
\newblock Full-wave analysis of thermal noise in antenna arrays on top of
  layered medium.
\newblock \emph{IEEE Transactions on Antennas and Propagation}, 72:\penalty0
  7560--7573, 2024.
\newblock \doi{10.1109/TAP.2024.3445132}.

\bibitem[Tutt et~al.(2026{\natexlab{b}})Tutt, Anstey, Cumner, Bevins, Cavillot,
  and de~Lera~Acedo]{Tutt_2026b}
Jacob~L. Tutt, Dominic~J. Anstey, John Cumner, Harry T.~J. Bevins, Jean
  Cavillot, and Eloy de~Lera~Acedo.
\newblock Towards end-to-end bayesian forward models in global 21-cm cosmology:
  surrogate modelling and marginalisation of beam uncertainty,
  2026{\natexlab{b}}.
\newblock URL \url{https://arxiv.org/abs/2608.18962}.

\bibitem[Pieterse et~al.(2024)Pieterse, Venter, and de~Villiers]{Pieterse_2024}
Carla~M. Pieterse, Mariet Venter, and Dirk I.~L. de~Villiers.
\newblock Investigating frequency dependency in characteristic basis function
  pattern modelling with geometric perturbations.
\newblock In \emph{2024 International Conference on Electromagnetics in
  Advanced Applications (ICEAA)}, pages 407--412, 2024.
\newblock \doi{10.1109/ICEAA61917.2024.10701964}.
\newblock URL \url{https://doi.org/10.1109/ICEAA61917.2024.10701964}.

\bibitem[Scheutwinkel et~al.(2022)Scheutwinkel, de~Lera~Acedo, and
  Handley]{Scheutwinkel_2022a}
K.~H. Scheutwinkel, E.~de~Lera~Acedo, and W.~Handley.
\newblock Bayesian evidence-driven diagnosis of instrumental systematics for
  sky-averaged 21-cm cosmology experiments.
\newblock \emph{Publications of the Astronomical Society of Australia},
  39:\penalty0 e052, 2022.
\newblock ISSN 1448-6083.
\newblock \doi{10.1017/pasa.2022.49}.
\newblock URL \url{http://dx.doi.org/10.1017/pasa.2022.49}.

\bibitem[Mittal et~al.(2024)Mittal, Kulkarni, Anstey, and
  de Lera Acedo]{Mittal_2024}
Shikhar Mittal, Girish Kulkarni, Dominic Anstey, and Eloy de Lera Acedo.
\newblock Impact of extragalactic point sources on the low-frequency sky
  spectrum and cosmic dawn global 21-cm measurements.
\newblock \emph{Monthly Notices of the Royal Astronomical Society},
  534\penalty0 (2):\penalty0 1317--1328, 09 2024.
\newblock ISSN 0035-8711.
\newblock \doi{10.1093/mnras/stae2111}.
\newblock URL \url{https://doi.org/10.1093/mnras/stae2111}.

\bibitem[Robins et~al.(2026)Robins, Anstey, Bevins, de~Lera~Acedo, and
  Irfan]{Robins_2026}
Daniel Robins, Dominic Anstey, Harry Bevins, Eloy de~Lera~Acedo, and Melis~O.
  Irfan.
\newblock Synchrotron and free--free mapping with simulated {REACH}
  observations between 50 and 170~{MHz}.
\newblock \emph{Monthly Notices of the Royal Astronomical Society},
  552\penalty0 (4):\penalty0 stag1804, 2026.
\newblock \doi{10.1093/mnras/stag1804}.
\newblock URL \url{https://doi.org/10.1093/mnras/stag1804}.

\bibitem[Shen et~al.(2021)Shen, Anstey, de Lera Acedo, Fialkov, and
  Handley]{Shen_2021}
Emma Shen, Dominic Anstey, Eloy de Lera Acedo, Anastasia Fialkov, and Will
  Handley.
\newblock Quantifying ionospheric effects on global 21-cm observations.
\newblock \emph{Monthly Notices of the Royal Astronomical Society},
  503\penalty0 (1):\penalty0 344--353, 02 2021.
\newblock ISSN 0035-8711.
\newblock \doi{10.1093/mnras/stab429}.
\newblock URL \url{https://doi.org/10.1093/mnras/stab429}.

\bibitem[Shen et~al.(2022)Shen, Anstey, de Lera Acedo, and
  Fialkov]{Shen_2022}
Emma Shen, Dominic Anstey, Eloy de Lera Acedo, and Anastasia Fialkov.
\newblock Bayesian data analysis for sky-averaged 21-cm experiments in the
  presence of ionospheric effects.
\newblock \emph{Monthly Notices of the Royal Astronomical Society},
  515\penalty0 (3):\penalty0 4565--4573, 07 2022.
\newblock ISSN 0035-8711.
\newblock \doi{10.1093/mnras/stac1900}.
\newblock URL \url{https://doi.org/10.1093/mnras/stac1900}.

\bibitem[Pattison et~al.(2023)Pattison, Anstey, and
  de Lera Acedo]{Pattison_2023}
Joe H~N Pattison, Dominic~J Anstey, and Eloy de Lera Acedo.
\newblock Modelling a hot horizon in global 21-cm experimental foregrounds.
\newblock \emph{Monthly Notices of the Royal Astronomical Society},
  527\penalty0 (2):\penalty0 2413--2425, 11 2023.
\newblock ISSN 0035-8711.
\newblock \doi{10.1093/mnras/stad3378}.
\newblock URL \url{https://doi.org/10.1093/mnras/stad3378}.

\bibitem[Pattison et~al.(2025)Pattison, Cavillot, Bevins, Anstey, Cumner, and
  de Lera Acedo]{pattison_2025a}
Joe H~N Pattison, Jean Cavillot, Harry T~J Bevins, Dominic~J Anstey, John~M
  Cumner, and Eloy de Lera Acedo.
\newblock Global 21 cm signal recovery under changing environmental
  conditions.
\newblock \emph{Monthly Notices of the Royal Astronomical Society},
  538\penalty0 (3):\penalty0 1301--1313, 02 2025.
\newblock ISSN 0035-8711.
\newblock \doi{10.1093/mnras/staf315}.
\newblock URL \url{https://doi.org/10.1093/mnras/staf315}.

\bibitem[Pattison et~al.(2026{\natexlab{b}})Pattison, Anstey, and
  de~Lera~Acedo]{pattison_2025b}
Joe H.~N. Pattison, Dominic~J. Anstey, and Eloy de~Lera~Acedo.
\newblock Quantifying the impact of lunar and planetary occultation on
  experimental global 21 cm cosmology.
\newblock \emph{Monthly Notices of the Royal Astronomical Society},
  546\penalty0 (4):\penalty0 stag258, 2026{\natexlab{b}}.
\newblock \doi{10.1093/mnras/stag258}.

\bibitem[Roque et~al.(2025)]{Roque_2025_receiver}
Ian L.~V. Roque et~al.
\newblock Receiver design for the {REACH} global 21-cm signal experiment.
\newblock \emph{Experimental Astronomy}, 59\penalty0 (1):\penalty0 7, 2025.
\newblock \doi{10.1007/s10686-024-09975-3}.
\newblock URL \url{https://doi.org/10.1007/s10686-024-09975-3}.

\bibitem[Kirkham et~al.(2025)Kirkham, Handley, Zhu, Artuc, Roque, Leeney,
  Bevins, Anstey, and de~Lera~Acedo]{Kirkham_2025}
Christian~J. Kirkham, William~J. Handley, Jiacong Zhu, Kaan Artuc, Ian L.~V.
  Roque, Samuel A.~K. Leeney, Harry T.~J. Bevins, Dominic~J. Anstey, and Eloy
  de~Lera~Acedo.
\newblock Accounting for noise and singularities in bayesian calibration
  methods for global 21-cm cosmology experiments.
\newblock \emph{Monthly Notices of the Royal Astronomical Society},
  543\penalty0 (4):\penalty0 4312--4320, 2025.
\newblock \doi{10.1093/mnras/staf1715}.
\newblock URL \url{https://doi.org/10.1093/mnras/staf1715}.

\bibitem[Handley et~al.(2015)Handley, Hobson, and Lasenby]{polychord}
W.~J. Handley, M.~P. Hobson, and A.~N. Lasenby.
\newblock polychord: next-generation nested sampling.
\newblock \emph{Monthly Notices of the Royal Astronomical Society},
  453\penalty0 (4):\penalty0 4384--4398, 09 2015.
\newblock ISSN 0035-8711.
\newblock \doi{10.1093/mnras/stv1911}.
\newblock URL \url{https://doi.org/10.1093/mnras/stv1911}.

\bibitem[Lueckmann et~al.(2021)Lueckmann, Boelts, Greenberg, Goncalves, and
  Macke]{Lueckmann_2021}
Jan-Matthis Lueckmann, Jan Boelts, David Greenberg, Pedro Goncalves, and Jakob
  Macke.
\newblock Benchmarking simulation-based inference.
\newblock In \emph{Proceedings of the 24th International Conference on
  Artificial Intelligence and Statistics}, volume 130 of \emph{Proceedings of
  Machine Learning Research}, pages 343--351. PMLR, 2021.
\newblock URL \url{https://proceedings.mlr.press/v130/lueckmann21a.html}.

\bibitem[Alvey et~al.(2024)Alvey, Bhardwaj, Domcke, Pieroni, and
  Weniger]{Alvey_2024_SGWB}
James Alvey, Uddipta Bhardwaj, Valerie Domcke, Mauro Pieroni, and Christoph
  Weniger.
\newblock Simulation-based inference for stochastic gravitational wave
  background data analysis.
\newblock \emph{Physical Review D}, 109\penalty0 (8):\penalty0 083008, 2024.
\newblock \doi{10.1103/PhysRevD.109.083008}.
\newblock URL \url{https://doi.org/10.1103/PhysRevD.109.083008}.

\bibitem[Gon{\c{c}}alves et~al.(2020)Gon{\c{c}}alves, Lueckmann, Deistler,
  Nonnenmacher, {\"O}cal, Bassetto, Chintaluri, Podlaski, Haddad, Vogels,
  Greenberg, and Macke]{Goncalves_2020}
Pedro~J. Gon{\c{c}}alves, Jan-Matthis Lueckmann, Michael Deistler, Marcel
  Nonnenmacher, Kaan {\"O}cal, Giacomo Bassetto, Chaitanya Chintaluri,
  William~F. Podlaski, Sara~A. Haddad, Tim~P. Vogels, David~S. Greenberg, and
  Jakob~H. Macke.
\newblock Training deep neural density estimators to identify mechanistic
  models of neural dynamics.
\newblock \emph{eLife}, 9:\penalty0 e56261, 2020.
\newblock \doi{10.7554/eLife.56261}.
\newblock URL \url{https://doi.org/10.7554/eLife.56261}.

\bibitem[Deistler et~al.(2025{\natexlab{a}})Deistler, Kadhim, Pals, Beck,
  Huang, Gloeckler, Lappalainen, Schr{"o}der, Berens, Gon{\c{c}}alves, and
  Macke]{Deistler_2025_Jaxley}
Michael Deistler, Kyra~L. Kadhim, Matthijs Pals, Jonas Beck, Ziwei Huang,
  Manuel Gloeckler, Janne~K. Lappalainen, Cornelius Schr{"o}der, Philipp
  Berens, Pedro~J. Gon{\c{c}}alves, and Jakob~H. Macke.
\newblock {Jaxley}: differentiable simulation enables large-scale training of
  detailed biophysical models of neural dynamics.
\newblock \emph{Nature Methods}, 22:\penalty0 2649--2657, 2025{\natexlab{a}}.
\newblock \doi{10.1038/s41592-025-02895-w}.
\newblock URL \url{https://doi.org/10.1038/s41592-025-02895-w}.

\bibitem[Tanoh et~al.(2025)Tanoh, Deistler, Macke, and Linderman]{Tanoh_2025}
Ian~Christopher Tanoh, Michael Deistler, Jakob~H. Macke, and Scott~W.
  Linderman.
\newblock Identifying multi-compartment {Hodgkin--Huxley} models with
  high-density extracellular voltage recordings.
\newblock In \emph{Advances in Neural Information Processing Systems},
  volume~38, 2025.
\newblock \doi{10.52202/085713-3307}.
\newblock URL
  \url{https://proceedings.neurips.cc/paper_files/paper/2025/hash/8e9c9bbe03d0da2d61e648e751ae61a4-Abstract-Conference.html}.

\bibitem[Yallup(2026)]{yallup2026nestedsamplingslicewithingibbsefficient}
David Yallup.
\newblock Nested sampling with slice-within-{Gibbs}: Efficient evidence
  calculation for hierarchical {Bayesian} models, 2026.
\newblock URL \url{https://arxiv.org/abs/2602.17414}.

\bibitem[Yallup et~al.(2025)Yallup, Prathaban, Alvey, and
  Handley]{yallup_2025_gw}
David Yallup, Metha Prathaban, James Alvey, and Will Handley.
\newblock Parallel nested slice sampling for gravitational wave parameter
  estimation, 2025.
\newblock URL \url{https://arxiv.org/abs/2509.24949}.

\bibitem[Deistler et~al.(2025{\natexlab{b}})Deistler, Boelts, Steinbach, Moss,
  Moreau, Gl{"o}ckler, Rodrigues, Linhart, Lappalainen, Miller,
  Gon{\c{c}}alves, Lueckmann, Schr{"o}der, and Macke]{Deistler_2025_SBI_Guide}
Michael Deistler, Jan Boelts, Peter Steinbach, Guy Moss, Thomas Moreau, Manuel
  Gl{"o}ckler, Pedro L.~C. Rodrigues, Julia Linhart, Janne~K. Lappalainen,
  Benjamin~Kurt Miller, Pedro~J. Gon{\c{c}}alves, Jan-Matthis Lueckmann,
  Cornelius Schr{"o}der, and Jakob~H. Macke.
\newblock Simulation-based inference: A practical guide, 2025{\natexlab{b}}.
\newblock URL \url{https://arxiv.org/abs/2508.12939}.

\bibitem[Gl{\"o}ckler et~al.(2022)Gl{\"o}ckler, Deistler, and
  Macke]{Glockler_2022}
Manuel Gl{\"o}ckler, Michael Deistler, and Jakob~H. Macke.
\newblock Variational methods for simulation-based inference.
\newblock In \emph{International Conference on Learning Representations}, 2022.
\newblock URL \url{https://openreview.net/forum?id=kZ0UYdhqkNY}.

\bibitem[Lyu et~al.(2026)Lyu, Alvey, Anau~Montel, Pieroni, and
  Weniger]{Lyu_2026_DynamicSBI}
Huifang Lyu, James Alvey, Noemi Anau~Montel, Mauro Pieroni, and Christoph
  Weniger.
\newblock Dynamic {SBI}: round-free sequential simulation-based inference with
  adaptive datasets.
\newblock \emph{Machine Learning: Science and Technology}, 7\penalty0
  (2):\penalty0 025036, 2026.
\newblock \doi{10.1088/2632-2153/ae51e1}.
\newblock URL \url{https://doi.org/10.1088/2632-2153/ae51e1}.

\bibitem[Alvey et~al.(2026)Alvey, Contaldi, and Pieroni]{Alvey_2026}
James Alvey, Carlo~R. Contaldi, and Mauro Pieroni.
\newblock Simulation-based inference with deep ensembles: evaluating
  calibration uncertainty and detecting model misspecification.
\newblock \emph{Machine Learning: Science and Technology}, 7\penalty0
  (1):\penalty0 015008, 2026.
\newblock \doi{10.1088/2632-2153/ae3103}.
\newblock URL \url{https://doi.org/10.1088/2632-2153/ae3103}.

\bibitem[Leeney et~al.(2026)Leeney, Gessey-Jones, Handley, de~Lera~Acedo,
  Bevins, and Tutt]{leeney_2026}
S.~A.~K. Leeney, T.~Gessey-Jones, W.~J. Handley, E.~de~Lera~Acedo, H.~T.~J.
  Bevins, and J.~L. Tutt.
\newblock Conditional neural bayes ratio estimation for experimental design
  optimisation, 2026.
\newblock URL \url{https://arxiv.org/abs/2603.26489}.

\bibitem[Cai et~al.(2024)Cai, Modi, Margossian, Gower, Blei, and
  Saul]{Cai_2024_EigenVI}
Diana Cai, Chirag Modi, Charles~C. Margossian, Robert~M. Gower, David~M. Blei,
  and Lawrence~K. Saul.
\newblock {EigenVI}: score-based variational inference with orthogonal function
  expansions.
\newblock In \emph{Advances in Neural Information Processing Systems},
  volume~37, 2024.
\newblock \doi{10.52202/079017-4218}.
\newblock URL
  \url{https://proceedings.neurips.cc/paper_files/paper/2024/hash/ef72fa6579401ffff9da246a5014f055-Abstract-Conference.html}.

\bibitem[Gunapati et~al.(2022)Gunapati, Jain, Srijith, and
  Desai]{Gunapati_2022}
Geetakrishnasai Gunapati, Anirudh Jain, P.~K. Srijith, and Shantanu Desai.
\newblock Variational inference as an alternative to {MCMC} for parameter
  estimation and model selection.
\newblock \emph{Publications of the Astronomical Society of Australia},
  39:\penalty0 e001, 2022.
\newblock ISSN 1448-6083.
\newblock \doi{10.1017/pasa.2021.64}.
\newblock URL \url{http://dx.doi.org/10.1017/pasa.2021.64}.

\bibitem[McIntosh-Smith et~al.(2024)McIntosh-Smith, Alam, and
  Woods]{Isamabrd_2024}
Simon McIntosh-Smith, Sadaf~R. Alam, and Christopher~J. Woods.
\newblock Isambard-{AI}: A leadership-class supercomputer optimised
  specifically for artificial intelligence.
\newblock In \emph{CUG 2024 Proceedings}, pages 44--54, 2024.

\bibitem[Pochinda et~al.(2024)Pochinda, Gessey-Jones, Bevins, Fialkov,
  Heimersheim, Abril-Cabezas, de~Lera~Acedo, Singh, Sikder, and
  Barkana]{Pochinda_2024}
Simon Pochinda, Thomas Gessey-Jones, Harry T.~J. Bevins, Anastasia Fialkov,
  Stefan Heimersheim, Irene Abril-Cabezas, Eloy de~Lera~Acedo, Saurabh Singh,
  Shikhar Sikder, and Rennan Barkana.
\newblock Constraining the properties of population iii galaxies with
  multi-wavelength observations.
\newblock \emph{Monthly Notices of the Royal Astronomical Society},
  531\penalty0 (1):\penalty0 1113--1132, 2024.
\newblock \doi{10.1093/mnras/stae1185}.

\bibitem[Dhandha et~al.(2025)Dhandha, Fialkov, Gessey-Jones, Bevins, Tacchella,
  Pochinda, de~Lera~Acedo, Singh, and Barkana]{Dhandha_2025}
Jiten Dhandha, Anastasia Fialkov, Thomas Gessey-Jones, Harry T.~J. Bevins,
  Sandro Tacchella, Simon Pochinda, Eloy de~Lera~Acedo, Saurabh Singh, and
  Rennan Barkana.
\newblock Exploiting synergies between {JWST} and cosmic 21-cm observations to
  uncover star formation in the early universe.
\newblock \emph{Monthly Notices of the Royal Astronomical Society},
  542\penalty0 (3):\penalty0 2292--2322, 2025.
\newblock \doi{10.1093/mnras/staf1359}.
\newblock URL \url{https://doi.org/10.1093/mnras/staf1359}.

\bibitem[{HERA Collaboration}(2022)]{HERA_2022_constraints}
{HERA Collaboration}.
\newblock {HERA} phase {I} limits on the cosmic 21-cm signal: Constraints on
  astrophysics and cosmology during the epoch of reionization.
\newblock \emph{The Astrophysical Journal}, 924\penalty0 (2):\penalty0 51,
  2022.
\newblock \doi{10.3847/1538-4357/ac2ffc}.

\bibitem[{HERA Collaboration}(2026)]{HERA_2026_phase2}
{HERA Collaboration}.
\newblock First results from {HERA} phase {II}.
\newblock \emph{The Astrophysical Journal}, 998\penalty0 (1):\penalty0 33,
  2026.
\newblock \doi{10.3847/1538-4357/ae2d54}.

\bibitem[{HERA Collaboration}(2023)]{HERA_2023_constraints}
{HERA Collaboration}.
\newblock Improved constraints on the 21 cm {EoR} power spectrum and the x-ray
  heating of the {IGM} with {HERA} phase {I} observations.
\newblock \emph{The Astrophysical Journal}, 945\penalty0 (2):\penalty0 124,
  2023.
\newblock \doi{10.3847/1538-4357/acaf50}.

\bibitem[Skilling(2006)]{Skilling_2006}
John Skilling.
\newblock {Nested sampling for general Bayesian computation}.
\newblock \emph{Bayesian Analysis}, 1\penalty0 (4):\penalty0 833--859, 2006.
\newblock \doi{10.1214/06-BA127}.

\end{thebibliography}

\newpage
\appendix

\section{Additional implementation and experimental details}
\label{app:supplementary_details}

\subsection{Astrophysical signal model}
\label{app:signal_physics}

\texttt{21cmSPACE} is a semi-numerical model that evolves a comoving simulation volume divided into $128^3$ cells, each with a volume of $(3~\mathrm{cMpc})^3$, from $z=50$ to the end of reionisation at $z=6$ \citep{Fialkov_2014,Cohen_2017}. Large-scale density and radiation fields are evolved explicitly, while halo and star formation are treated through sub-grid prescriptions. The resulting stellar populations source the Ly$\alpha$, X-ray, and ionising radiation that determines the coupling, temperature, and neutral fraction of the IGM, from which the sky-averaged brightness temperature $T_{21}(z)$ is calculated.

We adopt the inhomogeneous radio-galaxy extension introduced by \citet{Reis_2020_radio} and subsequently used in analyses of the first galaxies and HERA power-spectrum constraints \citep{Pochinda_2024,Dhandha_2025,HERA_2022_constraints,HERA_2026_phase2}. High-redshift galaxies are assigned a radio luminosity proportional to their star-formation rate,
\begin{equation}
\frac{L_{\mathrm{r}}(\nu)}{\mathrm{W\,Hz}^{-1}}
=
f_{\mathrm{radio}}10^{22}
\left(\frac{\nu}{150~\mathrm{MHz}}\right)^{-0.7}
\left(\frac{\mathrm{SFR}}{M_{\odot}\,\mathrm{yr}^{-1}}\right),
\label{eq:app_radio_luminosity}
\end{equation}
where $f_{\mathrm{radio}}=1$ corresponds to the luminosity of present-day radio galaxies. \texttt{21cmSPACE} propagates this spatially varying emission, modifying the radiation temperature against which the 21-cm transition is observed. Table~\ref{tab:app_signal_parameters} lists the nine quantities over which the simulation suite and emulator are defined.

\begin{table}[H]
    \centering
    \small
    \caption{Physical parameters and training ranges of the \texttt{21cmSPACE} simulations. The parameters $f_{\star,\mathrm{II}}$, $f_{\star,\mathrm{III}}$, $V_{\mathrm{c}}$, $f_{\mathrm{X}}$, and $f_{\mathrm{radio}}$ are sampled uniformly in their base-10 logarithm, $\tau$ is sampled uniformly in linear space, and $\alpha$, $\mathrm{pop}$, and $E_{\min}$ are evaluated on discrete grids.}
    \label{tab:app_signal_parameters}
    \begin{tabular}{@{}p{0.12\textwidth}p{0.46\textwidth}p{0.35\textwidth}@{}}
        \toprule
        Parameter & Physical role & Training range \\
        \midrule
        $f_{\star,\mathrm{II}}$ & Pop~II star-formation efficiency & $10^{-3}$--$10^{-0.3}$ \\
        $f_{\star,\mathrm{III}}$ & Pop~III star-formation efficiency & $10^{-3}$--$10^{-0.3}$ \\
        $V_{\mathrm{c}}$ & Minimum halo circular velocity for star formation & $4.2$--$100~\mathrm{km\,s^{-1}}$ \\
        $f_{\mathrm{X}}$ & X-ray luminosity efficiency & $10^{-3}$--$10^{3}$ \\
        $\tau$ & Thomson optical depth & $0.0235$--$0.1021$ \\
        $f_{\mathrm{radio}}$ & Radio luminosity relative to present-day galaxies & $10^{-1}$--$10^{5}$ \\
        $\alpha$ & Power-law slope of the X-ray SED & $\{1.0,1.3,1.5\}$ \\
        $\mathrm{pop}$ & Discrete Pop~III stellar-population prescription & $\{231,232,233\}$ \\
        $E_{\min}$ & Minimum escaping X-ray energy & $\{100,200,\ldots,1500,2000,3000\}~\mathrm{eV}$ \\
        \bottomrule
    \end{tabular}
\end{table}

\subsection{Neural-emulator dataset, training, and validation}
\label{app:signal_emulator}

The simulation ensemble contains 109,525 complete $T_{21}(z)$ realisations spanning the parameter space in Table~\ref{tab:app_signal_parameters}.\footnote{These simulations were precomputed for previous analyses \citep{Pochinda_2024,Dhandha_2025,HERA_2022_constraints,HERA_2023_constraints,HERA_2026_phase2}, at a total computational cost of $\sim9.20\times10^{5}$ CPU-core hours on Intel Xeon Platinum 8368Q processors with 6.76~GB of RAM per core.} Each simulation is evaluated on 200 redshift samples over $6\leq z\leq27$. We divide the complete simulations into training, validation, and test sets in a 60:20:20 ratio. This split is performed at the simulation level before redshift is folded into the network input, preventing samples from the same physical realisation from appearing in more than one subset.

Following the \texttt{globalemu} formulation \citep{Bevins_2021_globalemu}, redshift is treated as an additional input and the network learns the scalar mapping
\begin{equation}
\widehat{T}_{21}(z;\boldsymbol{\Theta})
=f_{\boldsymbol{\phi}}(z,\boldsymbol{\Theta}),
\qquad
\boldsymbol{\Theta}
=\{f_{\star,\mathrm{II}},f_{\star,\mathrm{III}},V_{\mathrm{c}},f_{\mathrm{X}},
\alpha,E_{\min},\tau,f_{\mathrm{radio}},\mathrm{pop}\}.
\label{eq:app_signal_emulator}
\end{equation}
The scale-spanning parameters $f_{\star,\mathrm{II}}$, $f_{\star,\mathrm{III}}$, $V_{\mathrm{c}}$, $f_{\mathrm{X}}$, and $f_{\mathrm{radio}}$ are transformed to base-10 logarithmic coordinates, while $\alpha$, $E_{\min}$, and $\mathrm{pop}$ enter as discrete conditioning variables drawn from their simulated grids. All inputs are standardised using the training set.

The emulator is a dense multilayer perceptron with four hidden layers of 64 units and GELU activations. We train against the simulated temperature in linear mK using a mean-squared-error objective and AdamW with a batch size of 1,000 and weight decay of $10^{-4}$. Training comprises 2,000 epochs: the learning rate decays exponentially from $10^{-3}$ to $5\times10^{-5}$ over the first 1,000 epochs, then from $5\times10^{-5}$ to $10^{-5}$ during a 1,000-epoch continuation. Early stopping is triggered at epoch 1,943 and this checkpoint is retained.\footnote{Training required approximately 73 minutes on a single NVIDIA H100 GPU.} Across the 21,905 held-out test spectra, the emulator achieves a mean per-signal RMSE of $0.91~\mathrm{mK}$.

For the benchmark presented here, we condition the emulator on $\alpha=1.3$, $E_{\min}=500~\mathrm{eV}$, and $\mathrm{pop}=232$. The directly inferred signal space is therefore
\begin{equation}
\boldsymbol{\theta}_{21}
=\{\log_{10}f_{\star,\mathrm{II}},\log_{10}f_{\star,\mathrm{III}},
\log_{10}V_{\mathrm{c}},\log_{10}f_{\mathrm{X}},\tau,
\log_{10}f_{\mathrm{radio}}\}.
\label{eq:app_signal_inference_parameters}
\end{equation}
Uniform priors over the emulator-training coordinates correspond to log-uniform priors on $f_{\star,\mathrm{II}}$, $f_{\star,\mathrm{III}}$, $V_{\mathrm{c}}$, $f_{\mathrm{X}}$, and $f_{\mathrm{radio}}$, and a uniform prior on $\tau$ in linear space.

\subsection{Foreground construction and priors}
\label{app:foreground_model}

To construct a full-fidelity foreground template that can be extrapolated across the observing band, we follow \citet{Anstey_2021} and evaluate the 2008 Global Sky Model (GSM) \citep{deOliveiraCosta2008} at 230 and 408~MHz. In celestial coordinates $\Omega'$, these maps define the spatially varying spectral-index field
\begin{equation}
\beta_{\mathrm{GSM}}(\Omega')
=
-\frac{
\log\!\left[
\bigl(T_{230}(\Omega')-T_{\mathrm{CMB}}\bigr)
/
\bigl(T_{408}(\Omega')-T_{\mathrm{CMB}}\bigr)
\right]
}{
\log(230/408)
},
\label{eq:app_spectral_index_map}
\end{equation}
where $T_{\mathrm{CMB}}=2.725~\mathrm{K}$. This field is then used to extrapolate the 230~MHz template to an arbitrary observing frequency:
\begin{equation}
T_{\mathrm{sky}}(\Omega',\nu)
=
\left[T_{230}(\Omega')-T_{\mathrm{CMB}}\right]
\left(\frac{\nu}{230~\mathrm{MHz}}\right)^{-\beta_{\mathrm{GSM}}(\Omega')}
+T_{\mathrm{CMB}}.
\label{eq:app_full_sky}
\end{equation}
This continuous pixel-level model is used to generate the benchmark data described in Appendix~\ref{app:data_generation}.

For inference, the spatial distributions of both the base-map temperature and spectral index are uncertain and cannot be treated as exact templates. A full pixel-level parameterisation would, however, be highly degenerate for the one- or two-dimensional data produced by a single radiometer. We therefore follow the regional construction of \citet{Anstey_2021,Anstey_2022} and \citet{Pagano_2023}, partitioning the sky independently into regions of uniform spectral index and base-map amplitude. In this work, we define $N_{\beta}=30$ spectral-index regions using equal percentiles of the full-resolution $\beta_{\mathrm{GSM}}$ field, such that each binary mask $M_{\beta,j}(\Omega')$ contains an equal fraction of the sky pixels. An equivalent set of masks $M_{\alpha,i}(\Omega')$ defines regions of uniform base-map correction. Their intersections give the two-dimensional masks used in Equation~\ref{eq:regional_sky_model}, $M_{i,j}(\Omega')=M_{\alpha,i}(\Omega')M_{\beta,j}(\Omega')$.

Each regional spectral-index parameter is assigned the same uniform prior spanning the minimum and maximum of the full spectral-index field,
\begin{equation}
\beta_j\sim\mathcal{U}(\beta_{\min},\beta_{\max}),
\qquad
\beta_{\min}=\min_{\Omega'}\beta_{\mathrm{GSM}}(\Omega'),
\quad
\beta_{\max}=\max_{\Omega'}\beta_{\mathrm{GSM}}(\Omega').
\label{eq:app_beta_prior}
\end{equation}
If uncertainty in the base-map amplitude is included, the corresponding multiplicative corrections are assigned
\begin{equation}
\alpha_i\sim\mathcal{N}(1,0.2^2),
\label{eq:app_alpha_prior}
\end{equation}
representing a $20\%$ uncertainty about the GSM template. The benchmark in this work isolates spectral-index and beam uncertainty and therefore uses a single amplitude region, $N_{\alpha}=1$, fixed to $\alpha_1=1$. The foreground consequently contributes 30 parameters to the sampled space.

\subsection{Beam-surrogate construction and validation}
\label{app:beam_surrogate}

We characterise each simulated beam using its normalised directivity, $D(\nu,\Omega)$, calculated from the orthogonal far-field electric-field components as
\begin{equation}
D(\nu,\Omega)
=
4\pi
\frac{|E_{\theta}(\nu,\Omega)|^2+|E_{\phi}(\nu,\Omega)|^2}
{\displaystyle\int_{4\pi}
\left[|E_{\theta}(\nu,\Omega)|^2+|E_{\phi}(\nu,\Omega)|^2\right]
\,\mathrm{d}\Omega}.
\label{eq:app_beam_directivity}
\end{equation}
The below-horizon response is set to zero before each frequency channel is renormalised such that $\int_{4\pi}D(\nu_i,\Omega)\,\mathrm{d}\Omega=4\pi$. We generate 500 beam realisations using the Method-of-Moments CEM solver described by \citet{Cavillot_2021,Cavillot2020,Cavillot2024}. Latin-hypercube sampling distributes these simulations across the seven physical perturbations listed in Table~\ref{tab:app_beam_parameters}. The ensemble is randomly divided into 400 training beams and 100 held-out validation beams.\footnote{Generating the complete ensemble required $8.0\times10^{5}$ CPU-core hours on Intel Xeon Platinum 8368Q processors with 6.76~GB of RAM per core.}

\begin{table}[H]
    \centering
    \caption{Physical parameters and ranges used to generate the CEM beam ensemble. Rotations and displacements are defined relative to the ground plane.}
    \label{tab:app_beam_parameters}
    \begin{tabular}{lll}
        \toprule
        Parameter & Symbol & Range \\
        \midrule
        Dipole blade height & $H_{\mathrm{blade}}$ & $800$--$1200~\mathrm{mm}$ \\
        Ground-plane height & $H_{\mathrm{GP}}$ & $1000$--$1500~\mathrm{mm}$ \\
        Dipole rotation & $\alpha_{\mathrm{rel}}$ & $-10$--$10~\mathrm{deg}$ \\
        Dipole $x$ displacement & $\Delta x_{\mathrm{rel}}$ & $-200$--$200~\mathrm{mm}$ \\
        Dipole $y$ displacement & $\Delta y_{\mathrm{rel}}$ & $-200$--$200~\mathrm{mm}$ \\
        Soil permittivity, real & $\epsilon_{\mathrm{real}}$ & $1$--$10$ \\
        Soil permittivity, imaginary & $\epsilon_{\mathrm{imag}}$ & $-2$--$0$ \\
        \bottomrule
    \end{tabular}
\end{table}

The first stage of the surrogate follows \citet{Pieterse_2024}. For each training beam $b$ and frequency channel $\nu_i$, the directivity is flattened over its $N_{\Omega}$ angular samples and stored as one row of the matrix $\mathbf{M}\in\mathbb{R}^{(N_{\mathrm{train}}N_{\nu})\times N_{\Omega}}$. Writing the combined beam--frequency row index as $\gamma=(b,i)$, its singular value decomposition is
\begin{equation}
M_{\gamma p}
=
\sum_{k=1}^{N_{\mathrm{mode}}}
U_{\gamma k}\,\sigma_k\,(V^{\dagger})_{kp},
\label{eq:app_beam_svd}
\end{equation}
where $p$ indexes angular samples, $N_{\mathrm{mode}}=\min(N_{\mathrm{train}}N_{\nu},N_{\Omega})$, and the singular values $\sigma_k$ are ordered by decreasing contribution to the beam variance. The rows of $\mathbf{V}^{\dagger}$ define the angular basis functions $B_k(\Omega)$, while the beam- and frequency-dependent weights are
\begin{equation}
w_{b,k}(\nu_i)=\sigma_k U_{(b,i),k}.
\label{eq:app_beam_svd_weights}
\end{equation}
A beam reconstructed with the first $R$ modes is therefore
\begin{equation}
\widetilde{D}_b^{(R)}(\nu_i,\Omega)
=
\sum_{k=1}^{R}w_{b,k}(\nu_i)B_k(\Omega).
\label{eq:app_beam_svd_reconstruction}
\end{equation}

The truncation rank is selected in antenna-temperature space, rather than from directivity error alone. Each truncated reconstruction is convolved with the full-resolution foreground sky for a conservative Galactic-centre-overhead snapshot,\footnote{Evaluated at 00:00:00 UTC on 2019-07-01 at the observing site ($\mathrm{lat}=-30.83875^\circ$, $\mathrm{lon}=21.37492^\circ$, altitude $1151~\mathrm{m}$), corresponding to an apparent LST of $20{:}00{:}31.86$.} and compared with the antenna temperature obtained using the corresponding full CEM beam. Retaining $R=400$ angular modes gives a mean RMSE of $20.09~\mathrm{mK}$ across the training ensemble, below the adopted $25~\mathrm{mK}$ instrumental-noise level. Directly parameterising the resulting frequency-dependent weights would nevertheless require $RN_{\nu}=34{,}000$ coefficients and would not encode their correlations across frequency and angular mode.

We therefore perform a second decomposition in the SVD-weight space. For each beam, the retained weights are stacked over frequency and angular mode into a vector $\mathbf{w}_b\in\mathbb{R}^{RN_{\nu}}$. After subtracting the training-set mean $\mathbf{e}_0$, an SVD of the residual beam-to-beam variations yields a new orthogonal set of correction vectors $\{\mathbf{e}_m\}$. In component form, the reconstructed weights are
\begin{equation}
\widetilde{w}_{b,k}^{(M)}(\nu_i)
=
e_{0,k}(\nu_i)
+
\sum_{m=1}^{M}a_{b,m}e_{m,k}(\nu_i),
\label{eq:app_pca_weight_reconstruction}
\end{equation}
which, when substituted into Equation~\ref{eq:app_beam_svd_reconstruction}, gives the surrogate in Equation~\ref{eq:beam_surrogate}. This second stage preserves the coherent variation of the beam across the observing band while reducing the instrumental representation to the coefficient vector $\boldsymbol{a}$.

The beam coefficients are centred by construction and assigned the Gaussian approximation
\begin{equation}
\boldsymbol{a}\sim
\mathcal{N}(\boldsymbol{0},\boldsymbol{\Sigma}_a),
\qquad
\boldsymbol{\Sigma}_a
=
\operatorname{diag}\!\left[
\operatorname{Var}(a_{b,1}),\ldots,
\operatorname{Var}(a_{b,M})
\right],
\label{eq:app_beam_coefficient_prior}
\end{equation}
where the variances are estimated across the 400 training beams. Applying the same antenna-temperature criterion after both decomposition stages shows that $M=100$ correction modes give a mean reconstruction RMSE of $24.47~\mathrm{mK}$. The final surrogate therefore compresses the original 34,000-dimensional SVD-weight space by a factor of 340. All benchmark observations use a full CEM beam from the held-out validation set, rather than a reconstruction used to train the surrogate (see \citet{Tutt_2026b} for details).

\subsection{Benchmark data generation}
\label{app:data_generation}

The benchmark datasets are generated independently of the hybrid representations used during inference. At each observing time, the full-resolution sky in Equation~\ref{eq:app_full_sky} is rotated from $\Omega'$ into the local antenna frame $\Omega$ and coupled to a full CEM validation beam, $D_{\mathrm{val}}(\nu,\Omega)$, which was excluded from the surrogate training set. For injected signal realisation $s$, the simulated data are
\begin{equation}
d_s(\nu,t)
=
\frac{1}{4\pi}
\int_{4\pi}
D_{\mathrm{val}}(\nu,\Omega)
T_{\mathrm{sky}}(\Omega,\nu,t)
\,\mathrm{d}\Omega
+T_{21}(\nu;\boldsymbol{\theta}_{21,s})
+\epsilon(\nu,t),
\label{eq:app_benchmark_data}
\end{equation}
where $\epsilon(\nu,t)\sim\mathcal{N}(0,\sigma_n^2)$ is uncorrelated Gaussian noise with $\sigma_n=25~\mathrm{mK}$.

We consider a three-hour observation from an observational site in the Karoo Desert, South Africa, beginning at 00:00:00 UTC on 2019-10-01. The observation is discretised into 1~MHz channels from 50 to 135~MHz and 5-minute time bins, giving $N_{\nu}=86$ and $N_t=37$. A test beam selected from the held-out set is used throughout, with $\{H_{\mathrm{blade}},H_{\mathrm{GP}},\alpha_{\mathrm{rel}},\Delta x_{\mathrm{rel}},\Delta y_{\mathrm{rel}},\epsilon_{\mathrm{real}},\epsilon_{\mathrm{imag}}\}=\{1080~\mathrm{mm},1272~\mathrm{mm},9.8^\circ,18~\mathrm{mm},-91~\mathrm{mm},4.24,-0.68\}$. The full-resolution foreground, beam, and observing window are fixed across the benchmark, isolating performance as the astrophysical signal changes. For further work investigating the robustness of the inference framework to different observing windows and beam relaisations see \citet{Tutt_2026a} and \citet{Tutt_2026b} respecitvely.

The 100 injected profiles are independent random draws from the six-dimensional prior defined in Appendix~\ref{app:signal_physics}. Consequently, the suite samples both strongly featured signals and broad, weakly identifiable profiles from the stated astrophysical model.

\subsection{Analytical beam marginalisation}
\label{app:beam_marginalisation}

For the marginalisation, the frequency--time data grid is flattened into a vector $\mathbf{d}\in\mathbb{R}^{N_{\mathrm{data}}}$, where $N_{\mathrm{data}}=N_{\nu}N_t$. With the amplitude corrections fixed to $\alpha_i=1$, we collect the sampled parameters as $\boldsymbol{\phi}=(\boldsymbol{\theta}_{\mathrm{FG}},\boldsymbol{\theta}_{21},\boldsymbol{\theta}_{\mathrm{Noise}})$. The mean-beam model is
\begin{equation}
\mu(\nu,t;\boldsymbol{\phi})
=
\sum_{j=1}^{N_{\beta}}
\mathcal{B}_{1,j,0}(\nu,t)
\left(\frac{\nu}{\nu_0}\right)^{-\beta_j}
+T_{\mathrm{CMB}}
+T_{21}(\nu;\boldsymbol{\theta}_{21}),
\label{eq:app_mean_beam_model}
\end{equation}
and the response of beam-correction mode $m$ is
\begin{equation}
h_m(\nu,t;\boldsymbol{\beta})
=
\sum_{j=1}^{N_{\beta}}
\mathcal{B}_{1,j,m}(\nu,t)
\left(\frac{\nu}{\nu_0}\right)^{-\beta_j}.
\label{eq:app_beam_correction_vector}
\end{equation}
Evaluating these expressions over the complete grid gives the vector $\boldsymbol{\mu}$ and design matrix
\begin{equation}
\mathbf{H}(\boldsymbol{\beta})
=
\left[\mathbf{h}_1\;\mathbf{h}_2\;\cdots\;\mathbf{h}_M\right],
\label{eq:app_beam_design_matrix}
\end{equation}
such that the forward-modelled data are linear in the beam coefficients,
\begin{equation}
\mathbf{d}
=
\boldsymbol{\mu}(\boldsymbol{\phi})
+\mathbf{H}(\boldsymbol{\beta})\boldsymbol{a}
+\boldsymbol{n}.
\label{eq:app_linear_beam_model}
\end{equation}

We assume the beam prior in Equation~\ref{eq:app_beam_coefficient_prior} and Gaussian noise $\boldsymbol{n}\sim\mathcal{N}(\boldsymbol{0},\mathbf{C}_n(\theta_n))$. The likelihood conditional on the beam coefficients is therefore
\begin{equation}
p(\mathbf{d}\mid\boldsymbol{\phi},\boldsymbol{a})
=
\mathcal{N}\!\left[
\mathbf{d}\,\middle|\,
\boldsymbol{\mu}(\boldsymbol{\phi})
+\mathbf{H}(\boldsymbol{\beta})\boldsymbol{a},
\mathbf{C}_n(\theta_n)
\right].
\label{eq:app_conditional_likelihood}
\end{equation}
Because both this likelihood and the prior on $\boldsymbol{a}$ are Gaussian, the instrumental parameters can be integrated out exactly:
\begin{equation}
\begin{split}
p(\mathbf{d}\mid\boldsymbol{\phi})
&=
\int
p(\mathbf{d}\mid\boldsymbol{\phi},\boldsymbol{a})
p(\boldsymbol{a})\,\mathrm{d}\boldsymbol{a}\\
&=
\mathcal{N}\!\left[
\mathbf{d}\,\middle|\,
\boldsymbol{\mu}(\boldsymbol{\phi}),
\mathbf{C}_{\mathrm{eff}}(\boldsymbol{\phi})
\right],
\end{split}
\label{eq:app_beam_integral}
\end{equation}
where
\begin{equation}
\mathbf{C}_{\mathrm{eff}}
=
\mathbf{C}_n
+\mathbf{H}\boldsymbol{\Sigma}_a\mathbf{H}^{\mathrm{T}}.
\label{eq:app_effective_covariance}
\end{equation}
Defining the residual $\mathbf{r}=\mathbf{d}-\boldsymbol{\mu}$, the marginal log-likelihood is
\begin{equation}
\ln\mathcal{L}_{\mathrm{marg}}
=
-\frac{1}{2}
\left[
\mathbf{r}^{\mathrm{T}}\mathbf{C}_{\mathrm{eff}}^{-1}\mathbf{r}
+\ln\det\mathbf{C}_{\mathrm{eff}}
+N_{\mathrm{data}}\ln(2\pi)
\right].
\label{eq:app_marginal_log_likelihood}
\end{equation}

Analytical marginalisation reduces the explicitly sampled model from 137 uncertain quantities to the 30 foreground, six astrophysical, and one noise parameter used by NSS, while retaining the covariance induced by the complete beam prior in both the posterior and evidence.

\subsection{Nested Slice Sampling configuration}
\label{app:nested_sampling}

We use Nested Sampling (NS) \citep{Skilling_2006} to obtain posterior samples and the Bayesian evidence,
\begin{equation}
\mathcal{Z}
=
\int \mathcal{L}(\boldsymbol{\phi})\pi(\boldsymbol{\phi})\,\mathrm{d}\boldsymbol{\phi}
=
\int_0^1 \mathcal{L}(X)\,\mathrm{d}X,
\label{eq:app_nested_evidence}
\end{equation}
where $X$ is the prior mass enclosed above a likelihood contour. Standard NS evolves a population of live points by repeatedly removing the lowest-likelihood point and replacing it with a draw satisfying the resulting likelihood constraint.

We use the GPU-native Nested Slice Sampling (NSS) implementation in \texttt{BlackJAX} \citep{yallup2025nested,cabezas2024blackjax}. As implemented by \citet{yallup2025nested}, each iteration removes the $b$ lowest-likelihood points and evolves $b$ independent replacement chains in parallel, all constrained above the highest likelihood among the removed batch. The expected contraction in log prior mass is
\begin{equation}
\mathbb{E}[\Delta\log X]
=
-\sum_{i=0}^{b-1}\frac{1}{n_{\mathrm{live}}-i}
\simeq
\log\!\left(\frac{n_{\mathrm{live}}-b}{n_{\mathrm{live}}}\right).
\label{eq:app_parallel_prior_contraction}
\end{equation}
Each replacement begins from a randomly selected surviving live point and is evolved using hit-and-run slice sampling within the likelihood-constrained region. Proposal directions are drawn in the Mahalanobis-normalised geometry defined by the covariance of the live population; stepping out and interval shrinkage then identify a valid point along each direction.

After analytical beam marginalisation, the sampled space has $D=37$ dimensions. All benchmark runs use $n_{\mathrm{live}}=25D=925$ live points, $n_{\mathrm{inner}}=8D=296$ slice steps per replacement, and $b=0.2n_{\mathrm{live}}=185$ simultaneous replacements. The live population is initialised independently from the prior for each injected dataset, and sampling terminates when the estimated remaining live-point evidence satisfies $\log\mathcal{Z}_{\mathrm{live}}-\log\mathcal{Z}<-3$.

\subsection{Signal-profile recovery diagnostics}
\label{app:recovery_metrics}

Recovery is evaluated through complementary diagnostics of posterior-mean accuracy and posterior calibration. For each benchmark realisation $r$, we systematically resample $S=2048$ equal-weight profiles, $T_r^{(s)}(\nu_k)$, from the weighted NSS posterior over $K$ frequency channels. For injected profile $T_r^\star(\nu_k)$, the posterior moments and RMSE are
\begin{equation}
\begin{aligned}
\overline{T}_{r,k}
&=\frac{1}{S}\sum_{s=1}^{S}T_r^{(s)}(\nu_k),
&
\widehat{\sigma}_{r,k}
&=\left[\frac{1}{S-1}\sum_{s=1}^{S}
\left(T_r^{(s)}(\nu_k)-\overline{T}_{r,k}\right)^2\right]^{1/2},\\
\operatorname{RMSE}_r
&=\left[\frac{1}{K}\sum_{k=1}^{K}
\left(T_r^\star(\nu_k)-\overline{T}_{r,k}\right)^2\right]^{1/2}.
\end{aligned}
\label{eq:app_profile_summary}
\end{equation}

RMSE quantifies point accuracy but not uncertainty calibration. We therefore assess coverage using the maximum marginally standardised deviation across frequency, a stringent whole-profile criterion requiring the credible band to contain every frequency channel:
\begin{equation}
\begin{aligned}
Z_r^\star
&=\max_k\left|
\frac{T_r^\star(\nu_k)-\overline{T}_{r,k}}{\widehat{\sigma}_{r,k}}
\right|,
&
Z_r^{(s)}
&=\max_k\left|
\frac{T_r^{(s)}(\nu_k)-\overline{T}_{r,k}}{\widehat{\sigma}_{r,k}}
\right|,\\
\mathcal{C}_{r,c}(\nu_k)
&=\left[
\overline{T}_{r,k}-q_{r,c}\widehat{\sigma}_{r,k},
\overline{T}_{r,k}+q_{r,c}\widehat{\sigma}_{r,k}
\right].
\end{aligned}
\label{eq:app_profile_band}
\end{equation}
where $q_{r,c}$ is the empirical $c$-quantile of $\{Z_r^{(s)}\}_{s=1}^{S}$. Estimating this quantile from complete posterior profiles retains their dependence across frequency. The minimum credible level required to contain the injection and the empirical coverage over $R$ realisations are
\begin{equation}
\begin{aligned}
\widehat{c}_r
&=\frac{1}{S}\sum_{s=1}^{S}
\left[
\mathbb{I}\!\left(Z_r^{(s)}<Z_r^\star\right)
+\frac{1}{2}\mathbb{I}\!\left(Z_r^{(s)}=Z_r^\star\right)
\right],\\
\widehat{\operatorname{Cov}}(c)
&=\frac{1}{R}\sum_{r=1}^{R}
\mathbb{I}\!\left(\widehat{c}_r\leq c\right).
\end{aligned}
\label{eq:app_profile_coverage}
\end{equation}
For $R=100$, $\widehat{\operatorname{Cov}}(0.95)=0.94$, consistent with the nominal $95\%$ level. Figure~\ref{fig:radio_galaxy_signal_recovery} is ordered by $\widehat{c}_r$ but displays pointwise intervals; coverage uses the whole-profile bands in Equation~\ref{eq:app_profile_band}.

\subsection{Astrophysical parameter posteriors}
\label{app:parameter_posteriors}

\begin{figure}[H]
    \centering
    \includegraphics[width=0.92\textwidth]{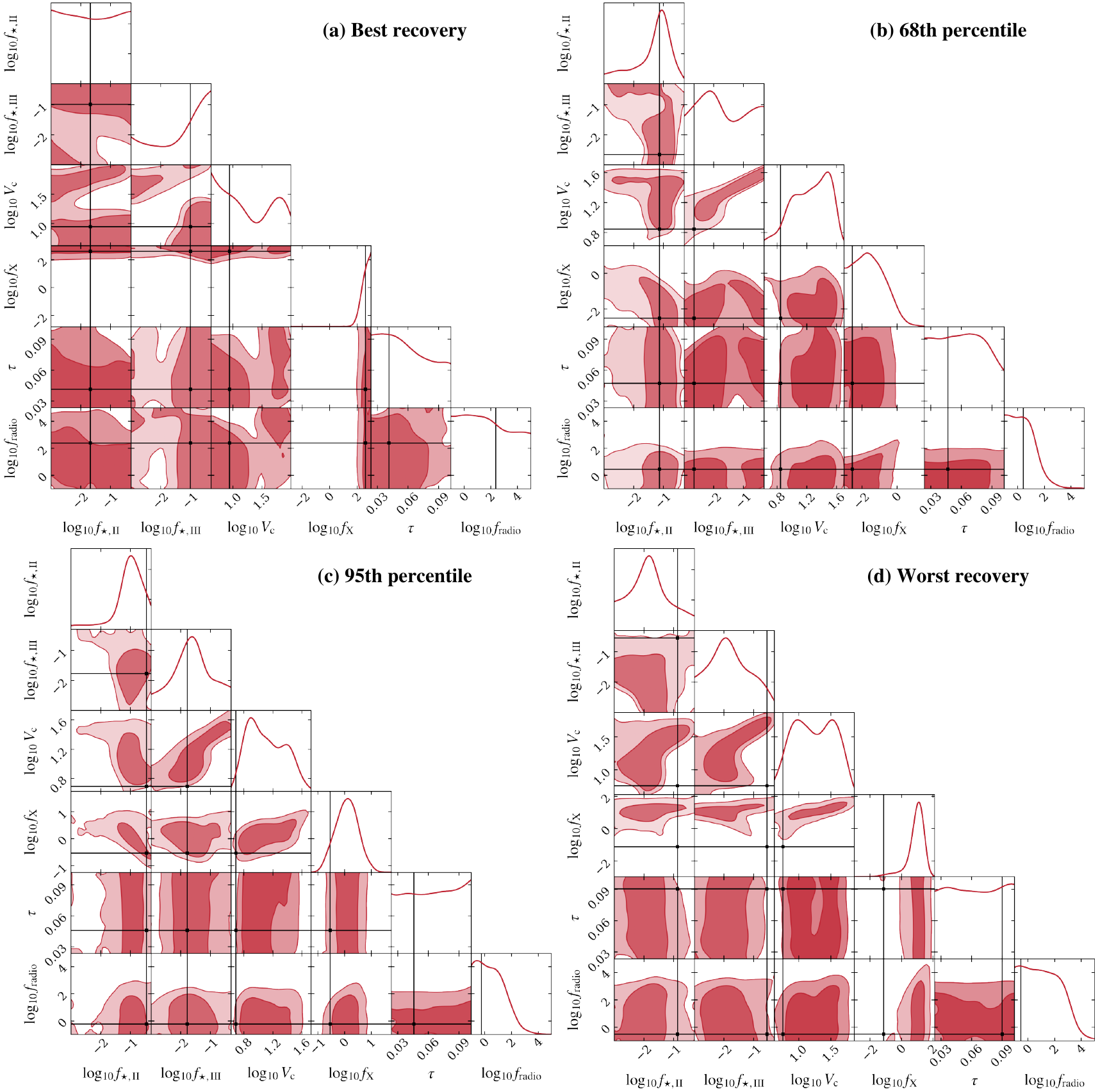}
    \caption{\textbf{Astrophysical parameter posteriors.} Marginal posteriors for the six signal parameters in the best, 50th-, 68th-, 95th-percentile, and worst cases ordered by minimum whole-profile credible level. Black lines and markers denote the injected parameters; dark and light red contours contain 68\% and 95\% posterior probability, respectively.}
    \label{fig:radio_galaxy_parameter_posteriors}
\end{figure}

To complement the signal-space recovery analysis in Section~\ref{sec:signal_recovery}, Figure~\ref{fig:radio_galaxy_parameter_posteriors} presents the corresponding six-dimensional astrophysical posteriors for the four representative cases shown in Figure~\ref{fig:radio_galaxy_signal_recovery}. These distributions illustrate how intrinsic parameter degeneracies and multimodality can map distinct astrophysical scenarios to similar global signal profiles.

\newpage
\section*{NeurIPS Paper Checklist}


\begin{enumerate}

\item {\bf Claims}
    \item[] Question: Do the main claims made in the abstract and introduction accurately reflect the paper's contributions and scope?
    \item[] Answer: \answerYes{}
    \item[] Justification: The abstract and introduction accurately state the paper's contributions and restrict the empirical claims to the observational configuration and uncertainty sources evaluated here. Broader observing windows and beam configurations have been evaluated in previous work and are discussed with the relevant citations.

\item {\bf Limitations}
    \item[] Question: Does the paper discuss the limitations of the work performed by the authors?
    \item[] Answer: \answerYes{}
    \item[] Justification: The paper provides a dedicated Limitations and Future Work section (Section~4), which discusses omitted observational and receiver-calibration effects, the increased posterior complexity expected from their inclusion, and the computational scaling limitations of the current sampling approach.

\item {\bf Theory assumptions and proofs}
    \item[] Question: For each theoretical result, does the paper provide the full set of assumptions and a complete (and correct) proof?
    \item[] Answer: \answerNA{}
    \item[] Justification: The paper does not claim new theorems or formal theoretical results. The underlying CEM and inhomogeneous radio-background simulators are established models described in the respective cited papers; the novel surrogate construction and analytical marginalisation used here are specified in full in the main text and appendix.

    \item {\bf Experimental result reproducibility}
    \item[] Question: Does the paper fully disclose all the information needed to reproduce the main experimental results of the paper to the extent that it affects the main claims and/or conclusions of the paper (regardless of whether the code and data are provided or not)?
    \item[] Answer: \answerYes{}
    \item[] Justification: The main text and appendix specify the forward-model components, parameter priors, data-generation procedure, emulator training, beam-surrogate construction, marginalised likelihood, sampling configuration, and recovery metrics required to reproduce the principal results.

\item {\bf Open access to data and code}
    \item[] Question: Does the paper provide open access to the data and code, with sufficient instructions to faithfully reproduce the main experimental results, as described in supplemental material?
    \item[] Answer: \answerYes{}
    \item[] Justification: The benchmark implementation and instructions are publicly available on \href{https://github.com/JacobTutt/global-21cm-inference-benchmark}{GitHub}. The synthetic benchmark datasets and inference products supporting the paper, including the NSS chains and production results, are publicly archived in \href{https://zenodo.org/records/22046858}{Zenodo record 22046858}. The original \texttt{21cmSPACE} training simulations and full CEM beam-generation and data-generation pipelines remain under private licences but are available upon reasonable request where permitted. These upstream assets are not required to rerun the released benchmark because the required emulator and response tensors are included.

\item {\bf Experimental setting/details}
    \item[] Question: Does the paper specify all the training and test details (e.g., data splits, hyperparameters, how they were chosen, type of optimizer) necessary to understand the results?
    \item[] Answer: \answerYes{}
    \item[] Justification: The main text and appendix report the data construction and splits, model architecture, optimiser, learning-rate schedule, training duration, parameter priors, beam-surrogate construction, and Nested Slice Sampling configuration used in the experiments.

\item {\bf Experiment statistical significance}
    \item[] Question: Does the paper report error bars suitably and correctly defined or other appropriate information about the statistical significance of the experiments?
    \item[] Answer: \answerYes{}
    \item[] Justification: Uncertainty is quantified directly by the Bayesian posterior distributions returned by Nested Slice Sampling. We report pointwise $95\%$ posterior credible bands for each recovered signal, while variation across experiments is summarised by the RMSE distribution and its percentiles over 100 held-out signal realisations.

\item {\bf Experiments compute resources}
    \item[] Question: For each experiment, does the paper provide sufficient information on the computer resources (type of compute workers, memory, time of execution) needed to reproduce the experiments?
    \item[] Answer: \answerYes{}
    \item[] Justification: The paper provides dedicated timing benchmarks and reports the CPU and GPU architectures, available memory, per-evaluation and per-run runtimes, training time, and aggregate compute required for the simulation ensembles and 100 inference runs.
    
\item {\bf Code of ethics}
    \item[] Question: Does the research conducted in the paper conform, in every respect, with the NeurIPS Code of Ethics \url{https://neurips.cc/public/EthicsGuidelines}?
    \item[] Answer: \answerYes{}
    \item[] Justification: The research uses simulated astrophysical, instrumental, and radiometer spectra and involves no personal data, human participants, surveillance, profiling, or consequential automated decision-making.

\item {\bf Broader impacts}
    \item[] Question: Does the paper discuss both potential positive societal impacts and negative societal impacts of the work performed?
    \item[] Answer: \answerYes{}
    \item[] Justification: By reducing simulation and inference costs, the framework broadens access to statistically rigorous uncertainty propagation, supports reproducible comparison of inference algorithms, and enables more reliable interpretation of global 21-cm observations. The work uses synthetic scientific data and is not expected to create material privacy, fairness, or dual-use risks.
    
\item {\bf Safeguards}
    \item[] Question: Does the paper describe safeguards that have been put in place for responsible release of data or models that have a high risk for misuse (e.g., pre-trained language models, image generators, or scraped datasets)?
    \item[] Answer: \answerNA{}
    \item[] Justification: The scientific simulator, synthetic spectra, and inference products contain no sensitive information and present no credible high-risk misuse pathway.

\item {\bf Licenses for existing assets}
    \item[] Question: Are the creators or original owners of assets (e.g., code, data, models), used in the paper, properly credited and are the license and terms of use explicitly mentioned and properly respected?
    \item[] Answer: \answerYes{}
    \item[] Justification: All existing software, simulation products, data, and models are correctly credited through the relevant citations and used with permission under their applicable terms. The benchmark code and documentation are released on \href{https://github.com/JacobTutt/global-21cm-inference-benchmark}{GitHub} under the BSD 3-Clause License, with the pretrained emulator weights provided under a separate 21cmSPACE Benchmark Model Use Licence. The benchmark datasets and inference products are archived on \href{https://zenodo.org/records/22046858}{Zenodo} under the Creative Commons Attribution 4.0 International (CC BY 4.0) licence.

\item {\bf New assets}
    \item[] Question: Are new assets introduced in the paper well documented and is the documentation provided alongside the assets?
    \item[] Answer: \answerYes{}
    \item[] Justification: The hybrid forward-model framework, synthetic benchmark datasets, and inference products introduced in this work are documented in the main text, appendix, and accompanying repository. The released materials describe their construction, parameterisation, provenance, structure, and limitations and are archived in the GitHub repository and Zenodo record cited above. No human-derived data are included.

\item {\bf Crowdsourcing and research with human subjects}
    \item[] Question: For crowdsourcing experiments and research with human subjects, does the paper include the full text of instructions given to participants and screenshots, if applicable, as well as details about compensation (if any)? 
    \item[] Answer: \answerNA{}
    \item[] Justification: No crowdsourcing, participant recruitment, or human-subject data collection was undertaken.

\item {\bf Institutional review board (IRB) approvals or equivalent for research with human subjects}
    \item[] Question: Does the paper describe potential risks incurred by study participants, whether such risks were disclosed to the subjects, and whether Institutional Review Board (IRB) approvals (or an equivalent approval/review based on the requirements of your country or institution) were obtained?
    \item[] Answer: \answerNA{}
    \item[] Justification: The study contains no human participants or identifiable human data and therefore does not require institutional human-subject review.

\item {\bf Declaration of LLM usage}
    \item[] Question: Does the paper describe the usage of LLMs if it is an important, original, or non-standard component of the core methods in this research? Note that if the LLM is used only for writing, editing, or formatting purposes and does \emph{not} impact the core methodology, scientific rigor, or originality of the research, declaration is not required.
    \item[] Answer: \answerNA{}
    \item[] Justification: LLMs were used only to refine text and assist with code debugging. They were not used in the simulator, model construction, inference methodology, data analysis, or experimental evaluation, and are not a component of the scientific method presented here.

\end{enumerate}

\end{document}